\documentclass{IEEE-Con-Sys-mag}

\usepackage[compress]{cite}
\usepackage{braket}
\usepackage{mathtools}
\usepackage{tikz}
\usetikzlibrary{decorations.pathreplacing}
\usetikzlibrary{shapes,arrows.meta}
\usetikzlibrary{calc}

\usepackage{times}

\usepackage{bbm}

\newcommand\rmd{\mathrm{d}}

\newcommand\rmx{\mathrm{x}}
\newcommand\rmy{\mathrm{y}}
\newcommand\rmz{\mathrm{z}}

\newcommand\rmG{\mathrm{G}}
\newcommand\rmH{\mathrm{H}}

\newcommand\rmU{\mathrm{U}}
\newcommand\rmV{\mathrm{V}}
\newcommand\rmW{\mathrm{W}}

\newcommand\bbc{\mathbb{C}}

\newcommand\bbr{\mathbb{R}}

\newcommand\bbu{\mathbb{U}}

\newcommand\calE{\mathcal{E}}
\newcommand\calF{\mathcal{F}}

\newcommand\calH{\mathcal{H}}

\newcommand\calM{\mathcal{M}}

\newcommand\calO{\mathcal{O}}

\newcommand\rmtr{\mathrm{tr}}

\jvol{XX}
\jnum{XX}
\paper{xx}
\jmonth{April}
\jname{submitted to IEEE CONTROL SYSTEMS}
\pubyear{2024}

\begin{document}

\title{Quantum computing through the lens of control\stitle{A tutorial introduction}}

\author{{J}ULIAN BERBERICH$^1$ and DANIEL FINK$^2$}
\affil{$^1$University of Stuttgart, Institute for Systems Theory and Automatic Control, 70569 Stuttgart, Germany.\\
$^2$University of Stuttgart, Institute for Computational Physics, 70569 Stuttgart, Germany.
}

\maketitle

\dois{}{}

\chapterinitial{Q}uantum computing is the science of storing and processing information using systems 
that obey the laws 
of quantum mechanics~\cite{nielsen2011quantum}.
Quantum mechanics describes nature on tiny scales, where it behaves radically different from our everyday experience.
On the atomic scale, systems exhibit 
counterintuitive effects such as entanglement 
(a strong form of coupling) or inherent and irresolvable uncertainty~\cite{feynman1965lectures3}.
It was first proposed in the 1980s by Richard Feynman that these effects could possibly be exploited 
to perform computations in a way that is superior to classical computing~\cite{feynman1982simulating}.

Soon after the first inception of quantum computers, algorithms have been developed which 
provably solve certain problems faster than any known classical algorithm.
For example, Grover's algorithm~\cite{grover1996fast} can be used to solve unstructured search problems over $N$ elements with a complexity 
of only 
$\calO(\sqrt{N})$.
Perhaps most prominently, Shor's algorithm~\cite{shor1999polynomial} allows to solve the integer factorization problem,
which is central to the RSA public-key encryption system, in polynomial time, that is, exponentially faster than the best known classical algorithm.
The simulation of quantum mechanical systems is 
another important application of quantum computing which may provide possible speedups over classic algorithms.
In particular, quantum simulation is inherently difficult for a classical computer and has, in fact, inspired the concept
of a quantum computer in the first place~\cite[Section 4.7]{nielsen2011quantum}, \cite{feynman1982simulating}.
These early theoretical successes have sparked a surge of research in the field and, in the meantime, quantum computing has evolved into a highly interdisciplinary
research field at the intersection of theoretical physics and computer science 
on the theoretical side, and experimental physics and engineering on the practical side.

\begin{summary}
    \summaryinitial{Q}uantum computing is a fascinating interdisciplinary research field that promises to revolutionize 
     computing by efficiently solving previously intractable problems.
 Recent years have seen tremendous progress on both the experimental realization of quantum computing devices as well as the 
 development and implementation of quantum algorithms. 
 Yet, realizing computational advantages of quantum computers in practice remains a widely open problem due to numerous fundamental challenges.
 Interestingly, many of these challenges are 
 connected to performance,
 robustness,
 scalability,
 optimization, or
 feedback,
 all of which are central concepts in control theory.
     This paper provides a tutorial introduction to quantum computing from the perspective of control theory.
     We introduce the mathematical framework of quantum algorithms ranging from basic elements including quantum bits and quantum gates to more advanced 
     concepts such as variational quantum algorithms and quantum errors.
     The tutorial only requires basic knowledge of linear algebra and, in particular, no prior exposure to quantum physics.    
     Our main goal is to equip readers with the mathematical basics required
      to understand and possibly solve (control-related) problems in quantum computing.
      In particular, beyond the tutorial introduction, we provide a list of
      research challenges in the field of quantum computing and discuss their connections to control.
    \end{summary}

Despite this substantial progress, realizing a computational advantage in practice on actual quantum computers
remains a central and widely open challenge.
In particular, current quantum computers have restricted capabilities:
The maximum number of possible \emph{qubits} (abbreviation of \emph{quantum bits}, which are the quantum analog of classical bits)
ranges from single-digit numbers over several dozens to few hundreds.
The maximum number of operations that can be applied is limited as well.
Finally, current quantum computers are strongly affected by various sources of noise, which can significantly perturb 
the outcome of a computation.
As a result, the current state of research in quantum computing 
is commonly referred to as the noisy intermediate-scale quantum (NISQ) era~\cite{preskill2018quantum,bharti2022noisy}.

In the NISQ era, implementing algorithms with theoretically proven speedups for meaningful problem sizes is beyond reach.
Instead, the focus has shifted towards studying the capabilities of available NISQ devices.
    For example, variational quantum algorithms (VQAs)~\cite{cerezo2021variational} are 
    a popular class of NISQ algorithms,
in which quantum computers are put into feedback with classical optimization schemes.
Recent years have seen impressive experimental progress of quantum computing on solving 
classically hard problems on quantum computers,
although mainly for specific problems of limited usage~\cite{arute2019quantum,kim2023evidence}.
Extending these first attempts to larger quantum computers with better reliability, solving more relevant problems,
and improving the understanding of quantum computers in general are the 
central goals of current research.

\section{Scope and structure of this tutorial}

This paper provides a tutorial introduction to quantum computing from the perspective 
of control theory (see ``Summary'').
We introduce the main algorithmic concepts, ranging from qubits and quantum gates as the central building blocks of quantum algorithms 
up to more advanced concepts such as VQAs and quantum errors.
The tutorial is written in a way that is accessible to readers without background in quantum physics and only requires 
basic knowledge of linear algebra as a prerequisite.
This is made possible via the mathematical framework of quantum computing that revolves around complex vectors and unitary matrices
but does not necessitate prior knowledge in quantum physics.

This tutorial has two main objectives.
First, it provides a basic introduction to the fascinating and active research field of quantum computing.
We cover the main mathematical concepts that can serve as a basis for following 
textbooks and research articles.
Throughout the paper, we mention several such follow-up references, most notably 
 the excellent textbook~\cite{nielsen2011quantum}.
Second, the goal of this tutorial is to promote the field of quantum computing as an important research branch that faces interesting challenges which 
are amenable to control techniques, including, for example, performance, robustness, scalability, optimization, or feedback.
In particular, we provide a list of research challenges in quantum computing and discuss their links to control.

Finally, let us comment on a popular research field at the intersection of quantum physics and control:
\emph{quantum control}.
The main goal of quantum control is the development and application of control methods for dynamical systems
obeying the laws of quantum mechanics~\cite{dong2010quantum,altafini2012modeling,dong2022quantum,koch2022quantum}.
These systems exhibit several unique properties which require the development of specialized control techniques.
It is important to emphasize that this tutorial is \emph{not} about quantum control.
Quantum control is mainly relevant for quantum computing in the experimental realization of a quantum computer,
which requires accurate control of microscopic quantities in real time.
Instead, this tutorial introduces the algorithmic framework of quantum computing, which allows to study and design 
quantum algorithms from an abstract mathematical viewpoint, independent of the physical hardware implementing the quantum computer.

This tutorial is structured as follows.
In the section ``Basic Elements of Quantum Computing'', we introduce qubits, measurement, and quantum gates, which are the basic building blocks of quantum computers.
We then combine these elements to form quantum algorithms, which are a combination of multiple qubits, series and parallel connections of quantum gates, and measurements  (Section ``Quantum Algorithms'').
Next, in the section ``Variational Quantum Algorithms'', we introduce the key concept and important examples of VQAs, which are among the most popular quantum algorithms in the recent literature.
From a control perspective, VQAs are particularly interesting since they are feedback interconnections of a discrete-time dynamical system with a static nonlinearity.
The section ``Density Matrices'' provides an alternative and 
often useful mathematical description of quantum algorithms.
Moreover, in the section ``Errors in Quantum Computing'', we introduce classes of errors occurring in quantum computing along with possibilities to mitigate and correct them.
The tutorial is concluded in the ``Conclusion'' section with a summary and discussion of the key concepts.

Tables~\ref{tab:abbreviations} and~\ref{tab:notation} introduce important abbreviations and notation used throughout this paper.

\begin{table}
    \caption{Abbreviations\label{tab:abbreviations}}
    \begin{tabular*}{18.5pc}{@{}p{40pt}p{180pt}<{\raggedright}@{}}
        CNOT:&controlled NOT\\
        NISQ:&noisy intermediate-scale quantum\\
        QAOA:&quantum approximate optimization algorithm\\
        QML:&quantum machine learning\\
        QEC:&quantum error correction\\
        QEM:&quantum error mitigation\\
        VQA:&variational quantum algorithm\\
        VQE:&variational quantum eigensolver\\
        ZNE:&zero-noise extrapolation
        \end{tabular*}        
    \end{table}

\begin{table}
    \caption{Notation\label{tab:notation}}
    \begin{tabular*}{18.5pc}{@{}p{40pt}p{180pt}<{\raggedright}@{}}
        $I_n$&This represents the $n\times n$-identity matrix.\\
        $\lVert x\rVert_p$&This represents the $p$-norm of a vector $x\in\bbc^n$.\\
        $A^\dagger$&This is the transposed and Hermitian conjugate of $A\in\bbc^{n\times n}$.\\
        $A\otimes B$&This represents the tensor (equivalently, Kronecker) product of $A,B\in\bbc^{n\times n}$.\\
        $\bbu^n$&This indicates the set of $n\times n$-dimensional unitary matrices, that is, matrices $U\in\bbc^{n\times n}$ satisfying $U^\dagger U=I$.\\
        $\ket{\psi}$&This is a quantum state $\psi\in\bbc^n$, $\lVert \psi\rVert_2=1$, (referred to as ``ket'').\\
        $\bra{\psi}$&This is the transposed and Hermitian conjugate of $\ket{\psi}$ (referred to as ``bra'').\\
        $\braket{\psi_1|\psi_2}$&This represents the inner product $\psi_1^\dagger\psi_2$ of two quantum states $\psi_1$, $\psi_2$.\\
        $\braket{\psi|H|\psi}$&This denotes the quadratic form $\psi^\dagger H\psi$ for a quantum state $\ket{\psi}\in\bbc^n$ and a matrix $H\in\bbc^{n\times n}$.\\
        $\ket{\psi_1\psi_2}$&This is a shorthand for $\ket{\psi_1}\otimes\ket{\psi_2}$.\\
        $\ket{\psi}^{\otimes n}$&This is a shorthand for $\underbrace{\ket{\psi}\otimes\cdots\otimes\ket{\psi}}_{n\>\text{times}}$.\\
        $\calH_1\otimes\calH_2$&This represents the tensor product of $\calH_1\subseteq\bbc^n$, $\calH_2\subseteq\bbc^n$.\\
        $\calH^{\otimes n}$&This is a shorthand for $\underbrace{\calH\otimes\cdots\otimes\calH}_{n\>\text{times}}$.
        \end{tabular*}        
    \end{table}

\section{Basic elements of quantum computing}\label{sec:basics_elements}

In the following, we introduce the basic ingredients of quantum computing.
We start by describing qubits, which are the main building blocks of quantum computers.
After explaining the extension to multiple qubits, we discuss the principle of measurement.
Finally, we present the concept of quantum gates along with some prominent examples.
The following exposition focuses on the key mathematical concepts as well as basic examples, and we refer to~\cite[Sections 1 \& 2]{nielsen2011quantum} for further details and additional insights.

\subsection{Qubits}\label{subsec:basics_qubit}
\emph{Qubits} are the basic unit in quantum computing, comparable to \emph{bits}
in classical computing.
A qubit is a two-level 
\emph{quantum state}, that is, a two-dimensional 
    complex vector
 $\ket{\psi}\in\bbc^2$ with unit norm $\lVert\ket{\psi}\rVert_2=1$.

 We use the standard \emph{Dirac notation} $\ket{\psi}$ for quantum states.
 For our purposes, $\ket{\psi}$ (called ``ket'') has the same mathematical meaning as 
 just writing $\psi$,
 and the notation can be viewed as a reminder that $\ket{\psi}$ is a quantum state.
 On the other hand, $\bra{\psi}$ (called ``bra'') is the transpose and Hermitian conjugate
 $\bra{\psi}\coloneqq\ket{\psi}^\dagger$.
 The terms \emph{bra} and \emph{ket} as well as their notation are motivated from the 
 inner product $\braket{\psi_1|\psi_2}=\psi_1^\dagger\psi_2$ (the ``bra-ket'' / ``bracket'').
 The notion of state in quantum physics is not to be confused with the notion of state used in control, compare~\cite{rosenbrock1999definition}.

Qubits are commonly represented in the standard basis
\begin{align}
    \ket0\coloneqq\begin{bmatrix}1\\0
\end{bmatrix},\>\ket1\coloneqq\begin{bmatrix}
    0\\1\end{bmatrix},
\end{align}
that is, there exist $\alpha,\beta\in\bbc$ such that
    \begin{align}\label{eq:qubit_basis}
        \ket{\psi}=\alpha\ket0+\beta\ket1
        =\begin{bmatrix}
            \alpha\\\beta
        \end{bmatrix}.
    \end{align}
    The states $\ket0$ and $\ket1$ are referred to as
    \emph{computational basis states}, and they play an analogous role to the two possible values $0$ and $1$
    of a classical bit.
    In contrast to classical computing, however, a qubit can lie in \emph{superposition}.
    This means that, in general, it is a linear combination of 
    the computational basis states as in~\eqref{eq:qubit_basis}.
    An important phenomenon that lies at the heart of quantum mechanics is that the precise value of 
    $\ket{\psi}$ (the values of $\alpha$ and $\beta$)
     cannot be measured directly.
    Instead, when measuring a qubit $\ket{\psi}$, the result is a classical bit, that is, there are only two possible outcomes
    $0$ or $1$.
    The probabilities for obtaining these outcomes are 
    $|\alpha|^2$ and $|\beta|^2$, respectively.
    Therefore, $\alpha$ and $\beta$ are called the \emph{probability amplitudes} of the state $\ket{\psi}$.
    The fact that $\ket{\psi}$ is a unit vector is consistent with this probabilistic interpretation since it 
    implies $|\alpha|^2+|\beta|^2=1$.
    Closely connected to this phenomenon is the \emph{collapse} of the qubit after measurement:
    If the measurement returns $0$ (or $1$), then, immediately after the measurement, the qubit state is equal to $\ket0$ (or $\ket1$).
    Later in the tutorial, a more rigorous introduction to measurements of quantum states is provided.

    Note that multiplication of a qubit in state $\ket{\psi}$ by a term $e^{-i\varphi}$ (called \emph{global phase}) with
    angle $\varphi\in\bbr$ does not affect the probability amplitudes:
    \begin{align}
        |e^{-i\varphi}\alpha|^2=|\alpha|^2,\quad
        |e^{-i\varphi}\beta|^2=|\beta|^2.
    \end{align}
    In particular, a global phase has no effect on the observable behavior 
    of a qubit, and two qubits which differ by a global phase 
    are considered equivalent.

\begin{figure}
    \centerline{
    \includegraphics[width=0.8\columnwidth]{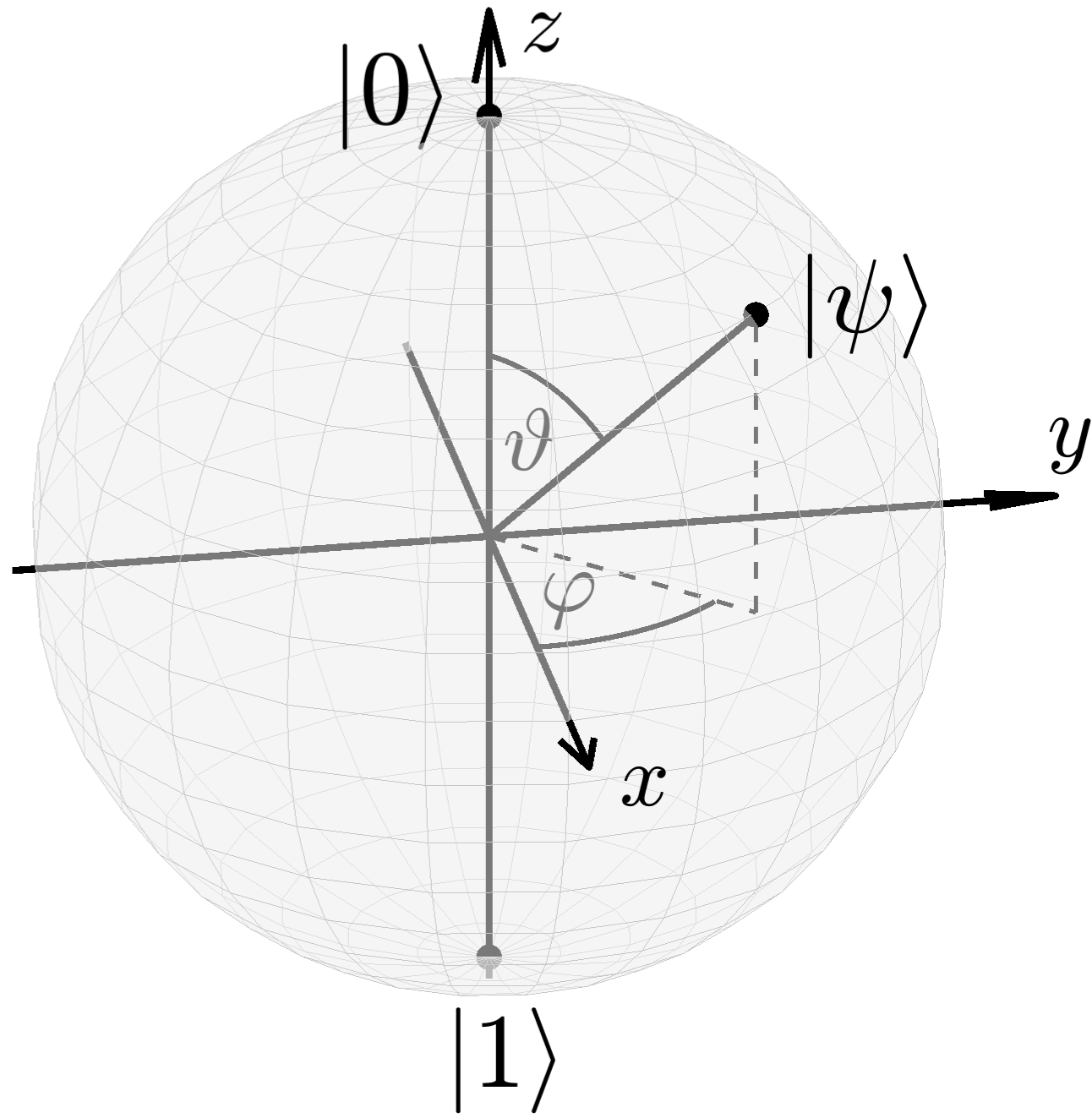}}
    \caption{Bloch sphere representation of a qubit $\ket{\psi}=\cos\frac{\vartheta}{2}\ket0+e^{i\varphi}\sin\frac{\vartheta}{2}\ket1$ for angles $\varphi$ and $\vartheta$.
    The Bloch sphere provides a useful illustration of a qubit, in particular how it generalizes a classical bit.
    While qubits can take values at any point on the sphere, classical bits can only take two values corresponding to $\ket0$ and $\ket1$.}
    \label{fig:bloch_sphere}
\end{figure}

    Recall that a qubit takes values in $\bbc^2$, which is isomorphic to $\bbr^4$, that is, described by four real parameters.
    However, the space of possible qubit values can be parameterized using only two real parameters:
    First, as explained above, multiplication by a global phase $e^{-i\varphi}$ does not change the qubit, which reduces the degrees of freedom by $1$.
    Second, qubits have unit norm, which further restricts the possible qubit values.
    In combination, a qubit can be equivalently represented using only two real parameters, that is,
    \begin{align}\label{eq:bloch_sphere_angles}
        \ket{\psi}=\cos\frac{\vartheta}{2}\ket0+e^{i\varphi}\sin\frac{\vartheta}{2}\ket1,
    \end{align}
    where $\vartheta,\varphi\in\bbr$ are the angles on the \emph{Bloch sphere} shown in Figure~\ref{fig:bloch_sphere},
    compare\ \cite[equation (1.4)]{nielsen2011quantum}.
The angles $\vartheta,\varphi$ can be derived from the representation~\eqref{eq:qubit_basis} based on elementary algebra.
To be precise, for a given qubit as in~\eqref{eq:qubit_basis} with $\alpha=r_{\alpha}e^{i\varphi_{\alpha}}$ and $\beta=r_{\beta}e^{i\varphi_{\beta}}$ for some $r_{\alpha},r_{\beta},\varphi_{\alpha},\varphi_{\beta}\geq0$, they can be computed as $\vartheta=2\cos^{-1}(r_{\alpha})$ and $\varphi=\varphi_{\beta}-\varphi_{\alpha}$.

The Bloch sphere is a useful illustration of single qubits, which also shows how they generalize classical bits:
While a classical bit can only take two values, which are represented in Figure~\ref{fig:bloch_sphere} as the North Pole $\ket0$ and the
South Pole $\ket1$, a qubit can lie anywhere on the Bloch sphere.
Thus, qubits have a significantly larger range of possible values, which already gives a first hint at the power of quantum computing.
Measurements of the qubit, however, can only result in either $\ket0$ or $\ket1$, where the precise position on the Bloch sphere
determines the respective probabilities.
Therefore, the main challenge in quantum computing is to exploit the range of possible qubit values (infinitely many) 
even when measurements collapse the qubit onto one of two discrete values.

The mathematical definition of a qubit given above can take different physical realizations.
In order to build a qubit in the real world, any quantum mechanical system which 
can take one of 
two values upon measurement
will do.
As an instructive example, a qubit can be implemented via the spin of an electron.
Whenever measuring the spin, it will be either \emph{up} or \emph{down}.
Between measurements, the spin evolves in a complex superposition of these two outcomes,
with the coefficients characterizing the measurement probabilities.
There are many other possible realizations of a qubit, and we briefly touch upon this point later in ``Programming and Experimental Realization of Quantum Algorithms''.
The beauty of quantum computing is that it allows for an abstract theoretical framework
based on complex vectors and unitary matrices, which encompasses a wide range of different 
physical realizations.
To some extent, this allows to separate the analysis and design of quantum algorithms from their physical implementation.

\begin{sidebar}{An Example of Entangled States: The Bell States}

    \setcounter{sequation}{0}
    \renewcommand{\thesequation}{S\arabic{sequation}}
    \setcounter{stable}{0}
    \renewcommand{\thestable}{S\arabic{stable}}
    \setcounter{sfigure}{0}
    \renewcommand{\thesfigure}{S\arabic{sfigure}}

    \sdbarinitial{T}he simplest example for entangled states are the \emph{Bell states}
        \begin{sequation}\label{eq:Bell_states}
            \ket{\Phi^+}=\frac{1}{\sqrt{2}}(\ket{00}+\ket{11}),
        \end{sequation}
        \begin{sequation}
            \ket{\Phi^-}=\frac{1}{\sqrt{2}}(\ket{00}-\ket{11}),
        \end{sequation}
    \begin{sequation}
            \ket{\Psi^+}=\frac{1}{\sqrt{2}}(\ket{01}+\ket{10}),
        \end{sequation}
        \begin{sequation}
            \ket{\Psi^-}=\frac{1}{\sqrt{2}}(\ket{01}-\ket{10}).
        \end{sequation}
        Indeed, it is straightforward to show that neither of these states can be factored into two single-qubit states as in~\eqref{eq:separable}.
        Intuitively, entanglement can be understood as follows:
        The Bell state $\ket{\Phi^+}$ is an equal superposition of the two computational basis states $\ket{00}$ and $\ket{11}$.
        By definition, it is equal to
        \begin{sequation}
            \ket{\Phi^+}=\frac{1}{\sqrt{2}}\begin{bmatrix}
                1\\0\\0\\1
            \end{bmatrix}.
        \end{sequation}
        Suppose now that we only measure the first qubit of $\ket{\Phi^+}$.
        If we obtain the result $0$, then we are guaranteed that the state of the first qubit is $\ket0$.
        Given the definition of $\ket{\Phi^+}$, this means that also the second qubit must be in state $\ket0$ (the probability amplitude corresponding to $\ket{01}$ is zero).
        Thus, measuring the first qubit uniquely determines the state of the second qubit, without taking another measurement!
        On the other hand, if we obtain the result $1$ for the first qubit, then the second qubit must be in state $\ket1$.
        This is the case even when the two qubits are physically separated by a large distance, which is why Einstein referred to entanglement as ``spooky action at a distance''.   
    \end{sidebar}

\subsection{Multiple qubits}\label{subsec:basics_multi_qubits}

Just like classical computers operate on many classical bits, quantum computers generally operate on multiple qubits.
We have seen in the previous section that qubits live in a subset of $\bbc^2$, which is isomorphic to a sphere in $\bbr^3$.
Suppose now that we have $n$ such qubits.
The quantum state $\ket{\psi}$ representing all $n$ qubits lives in the $n$-fold \emph{tensor product} of $\bbc^2$ with itself.
More precisely,
$\ket{\psi}$ is a unit vector in
\begin{align}\label{eq:composite_state_Hilbert_space}
   \big(\bbc^{2}\big)^{\otimes n}\coloneqq\underbrace{\bbc^2\otimes\cdots\otimes\bbc^2}_{n\>\text{times}}=\bbc^{2^n}.
\end{align}
A basis for $\bbc^{2^n}$ can be constructed by taking tensor products of the computational basis states $\ket0$ and $\ket1$, that is,
\begin{align}\label{eq:composite_state_computational_basis}
    \{\ket{0\dots000},\ket{0\dots001},\ket{0\dots010},\dots,\ket{1\dots111}\},
\end{align}
where we use the common notation
\begin{align}\label{eq:tensor_product_notation}
    \ket{\psi_1\psi_2\dots\psi_n}=\ket{\psi_1}\otimes\ket{\psi_2}\dots\ket{\psi_n}
\end{align}
for quantum states $\ket{\psi_j}$.
For example,
\begin{align*}
    \ket{0\dots010}=\ket0\otimes\dots\otimes\ket0\otimes\ket1\otimes\ket0.
\end{align*}
The basis~\eqref{eq:composite_state_computational_basis} is commonly referred to as the \emph{computational basis}, and its components are also frequently converted from their binary representation into decimal form, for example, $\ket{5}\coloneqq\ket{101}$.

For a given multi-qubit state $\ket{\psi}$, it may or may not be possible to decompose it as
\begin{align}\label{eq:separable}
    \ket{\psi}=\ket{\psi_1}\otimes\ket{\psi_2}
\end{align}
for states $\ket{\psi_1}$ and $\ket{\psi_2}$ corresponding to disjoint sets of qubits.
All states $\ket{\psi}$ which can be written as in~\eqref{eq:separable} are called \emph{separable}.
It is important to emphasize that \emph{not all states are separable}, that is, there are unit vectors in $\bbc^{2^n}$ which cannot be decomposed as in~\eqref{eq:separable}.
All the states that are not separable are called \emph{entangled}.
Entanglement is a mysterious property of quantum objects, describing a strong coupling between them which has no classical analog.
A sufficient degree of entanglement is necessary for quantum computing to achieve an exponential speedup over classical computing~\cite{jozsa2003role}.
In ``An Example of Entangled States: The Bell States'', we provide an example of entangled quantum states.

Let us now consider the extreme case of an $n$-qubit state consisting of $n$ individual qubits with corresponding states $\ket{\psi_j}$, $j=1,\dots,n$.
Mathematically, the composite state $\ket{\psi}$ is given by the tensor product of the $\ket{\psi_j}$'s, that is,
\begin{align}\label{eq:product_state}
    \ket{\psi}=\ket{\psi_1}\otimes\ket{\psi_2}\otimes\dots\otimes\ket{\psi_n}.
\end{align}
States of the form~\eqref{eq:product_state} are commonly referred to as \emph{product states}.
As a simple example, consider a two-qubit state consisting of two independent single-qubit states $\ket{\psi_1}=\begin{bmatrix}
    \alpha_1\\\beta_1
\end{bmatrix}$ and $\ket{\psi_2}=\begin{bmatrix}
    \alpha_2\\\beta_2
\end{bmatrix}$.
Note that, when applied to vectors or matrices as in~\eqref{eq:tensor_product_notation},
the tensor product is equivalent to an operation that is frequently used in control theory:
the \emph{Kronecker product}.
Hence, 
$\ket{\psi}$ is given by
\begin{align}
    \ket{\psi}&=\ket{\psi_1}\otimes\ket{\psi_2}
    =\begin{bmatrix}
        \alpha_1\alpha_2\\\alpha_1\beta_2\\\beta_1\alpha_2\\\beta_1\beta_2
    \end{bmatrix}\\\nonumber
    &=\alpha_1\alpha_2\ket{00}
    +\alpha_1\beta_2\ket{01}+\beta_1\alpha_2\ket{10}
    +\beta_1\beta_2\ket{11}.
\end{align}
Intuitively, this construction can be explained via the probabilistic interpretation of the amplitudes $\alpha_i$, $\beta_i$.
If the probability of measuring $0$ for the state $\ket{\psi_1}$, respectively $\ket{\psi_2}$, is $|\alpha_1|^2$, respectively $|\alpha_2|^2$,
then the probability of measuring $00$ for the combined state $\ket{\psi_1}\otimes\ket{\psi_2}$ is given by $|\alpha_1\alpha_2|^2$ (and similarly for the other possible measurement outcomes $01$, $10$, and $11$).

Let us conclude by emphasizing that the size of $\bbc^{2^n}$, the space of quantum states consisting of $n$ qubits, is exponentially large in $n$.
This shows that, in general, a quantum state can only be represented on a classical computer for a small number of qubits $n$, that is, for values of $n$ such that $2^n$ is not too large.

\begin{sidebar}{Projective Measurement as Expectation Estimation}

    \setcounter{sequation}{5}
    \renewcommand{\thesequation}{S\arabic{sequation}}
    \setcounter{stable}{0}
    \renewcommand{\thestable}{S\arabic{stable}}
    \setcounter{sfigure}{0}
    \renewcommand{\thesfigure}{S\arabic{sfigure}}

    \sdbarinitial{M}easurement of a quantum state $\ket{\psi}$ with respect to the observable $\calM$ can be understood as evaluating the quadratic form
\begin{sequation}\label{eq:measurement_quadratic_function}
    \braket{\psi|\calM|\psi}.
\end{sequation}
This value can be interpreted as the expectation of the observable for the given quantum state.
In what follows, we explain how it can be determined from the projective measurements explained above.
To this end, we use~\eqref{eq:measurement_projections_decomposition} to rewrite~\eqref{eq:measurement_quadratic_function} as 
\begin{sequation}
    \braket{\psi|\calM|\psi}= 
    \braket{\psi|\sum_{i=1}^{\ell}\lambda_iP_i|\psi}
    =\sum_{i=1}^{\ell}\lambda_i\braket{\psi|P_i|\psi}.    
\end{sequation}
That is, $\braket{\psi|\calM|\psi}$ can be computed as the sum over all products of measurement outcomes $\lambda_i$ with the corresponding 
probabilities $\braket{\psi|P_i|\psi}$.
This leads to a simple procedure for estimating $\braket{\psi|\calM|\psi}$:
Suppose we have access to $T$ copies of the state $\ket{\psi}$ and we perform measurements of each state with respect to $\calM$.
Multiple copies of $\ket{\psi}$ are required in order 
to perform multiple measurements of $\ket{\psi}$ since each measurement unavoidably influences $\ket{\psi}$.
If, for the given $T$ copies of $\ket{\psi}$, each $\lambda_i$ is measured $T_i$ times, then, for $T$ sufficiently large, we have 
\begin{sequation}\label{eq:measurement_quadratic_function_estimate}
    \braket{\psi|\calM|\psi}\approx\frac{\sum_{i=1}^{\ell}\lambda_iT_i}{T}.
\end{sequation}
While this provides a useful approximation of the quadratic form~\eqref{eq:measurement_quadratic_function}, it is 
also possible to make the argument more rigorous and study, for example, the estimation error
 (see, for example,~\cite[Section 3.2.4]{schuld2021machine}).

In practice, the state $\ket{\psi}$ which is supposed to be measured is often the outcome of a quantum algorithm.
In this case, producing $T$ copies of $\ket{\psi}$ means executing $T$ instants (referred to as \emph{shots}) of the same quantum algorithm.
\end{sidebar}

\subsection{Measurement}\label{subsec:basics_measurement}

As mentioned above, it is not possible to access the value of a quantum state directly.
Instead, we need to take measurements according to the laws of quantum mechanics.
In this tutorial, we focus on \emph{projective measurements} for simplicity, but we note that generalizations and variations do exist (compare~\cite[Section 2.2.6]{nielsen2011quantum}).
Projective measurements are always taken with respect to an \emph{observable} $\calM$, which is a Hermitian matrix of dimension $\ell=2^n$
 (for an $n$-qubit system), that is, 
$\calM=\calM^\dagger\in\bbc^{2^n}$.
By the spectral theorem, we can write 
\begin{align}\label{eq:measurement_projections_decomposition}
    \calM=\sum_{i=1}^{\ell}\lambda_iP_i,
\end{align}
where $\lambda_i\in\bbr$ are the eigenvalues of $\calM$ and $P_i$ are the projectors onto the corresponding eigenspaces, that is, 
$P_i=v_iv_i^\dagger$ with the eigenvectors $v_i$.
The outcome of a measurement is always one of the eigenvalues $\lambda_i$ of $\calM$, and the probability 
for measuring $\lambda_i$ is equal to
\begin{align}\label{eq:measurement_probability}
    \braket{\psi|P_i|\psi}=\psi^\dagger P_i\psi.
\end{align}
If the result of the measurement is given by $\lambda_i$, then, directly after the measurement, the state is equal to
\begin{align}\label{eq:measurement_collapse}
    \ket{\psi_{\mathrm{after}\>\mathrm{meas.}}}
    =\frac{P_i\ket{\psi}}{\sqrt{\braket{\psi|P_i|\psi}}},
\end{align}
compare\ \cite[Section 2.2.5]{nielsen2011quantum} for details.
Let us emphasize this rather peculiar fact:
The measurement \emph{influences} the quantum state $\ket{\psi}$.
At the time of the measurement, $\ket{\psi}$ changes its value and, at an infinitesimally short time \emph{after} the measurement, it is equal to one of the eigenvectors of the observable $\calM$.
This process is referred to as the \emph{collapse} of the quantum state $\ket{\psi}$.
The only information that is extracted via the measurement is one of the eigenvalues of $\calM$, which reveals onto which eigenvector the state has collapsed.
The fact that measurements affect the values of quantum states (possibly in an undesirable way) is, of course, bad news for feedback control.
In ``Projective Measurement as Expectation Estimation'', we introduce another frequently employed viewpoint on projective measurements as statistical estimates of a quadratic form.

    Let us illustrate the above principles with a single-qubit example.
        The most frequently considered observable is the \emph{Pauli matrix}
    $Z=\begin{bmatrix}
        1&0\\0&-1
    \end{bmatrix}$.
    Since the eigenvectors of $Z$ are the computational basis states $\ket0$ and $\ket1$, measurements with respect to $Z$ are commonly referred to as measurements in the computational basis.
    To provide the explicit formulas for such measurements, we use the spectral decomposition
    \begin{align}
        Z=\underbrace{1}_{\lambda_1=}\cdot
        \underbrace{\begin{bmatrix}
            1&0\\0&0
        \end{bmatrix}}_{P_1=}
        +\underbrace{(-1)}_{\lambda_2=}\cdot
        \underbrace{\begin{bmatrix}
            0&0\\0&1
        \end{bmatrix}}_{P_2=}.
    \end{align}
    Hence, measuring $Z$ for a single qubit $\ket{\psi}=\alpha\ket0+\beta\ket1$
    always yields one of the eigenvalues $\lambda_1=+1$ or $\lambda_2=-1$ of $Z$.
    According to~\eqref{eq:measurement_probability}, the probability for measuring $\lambda_1=+1$ is
    \begin{align}
        \braket{\psi|P_1|\psi}=\begin{bmatrix}
            \alpha\\\beta
        \end{bmatrix}^\dagger \begin{bmatrix}
            1&0\\0&0
        \end{bmatrix}
        \begin{bmatrix}
            \alpha\\\beta
        \end{bmatrix}=|\alpha|^2.
    \end{align}
    Combining this with~\eqref{eq:measurement_collapse}, if the measurement returns the value $\lambda_1=+1$, 
    then we can be certain that the qubit is in the state 
    \begin{align}\label{eq:measurement_example_Z}
        \frac{P_1\ket\psi}{\braket{\psi|P_1|\psi}}
=\frac{\begin{bmatrix}1&0\\0&0\end{bmatrix}
\begin{bmatrix}
    \alpha\\\beta
\end{bmatrix}}{\sqrt{|\alpha|^2}}
=e^{-\varphi_{\alpha}}\ket0
    \end{align}
with $\varphi_{\alpha}\in\bbr$ such that $e^{-\varphi_{\alpha}}=\frac{\alpha}{|\alpha|}$.
Recall that global phases $e^{-i\varphi_{\alpha}}$ do not influence measurement outcomes
(a fact that can now be seen from~\eqref{eq:measurement_probability} and~\eqref{eq:measurement_collapse} by multiplying $\psi$ by $e^{-i\varphi_{\alpha}}$).
Therefore, the state in~\eqref{eq:measurement_example_Z} is equivalent to $\ket0$.
To summarize, the probability for obtaining $+1$ as the measurement outcome is $|\alpha|^2$ and, if we measure $+1$, 
then the state collapses to $\ket0$ after the measurement.
Similarly, it can be shown that $|\beta|^2$ describes the probability for measuring $-1$, and the state collapses to $\ket1$ if
$-1$ is measured.

Finally, we note that, for many practical purposes, it is relevant to 
determine the full quantum state, that is, all its coefficients in the computational basis~\eqref{eq:composite_state_computational_basis}, rather than just a measurement with respect to a given observable.
Since the qubits are not directly accessible via measurements, different methods for doing so have been 
developed under the name of \emph{quantum state tomography} (see, for example,~\cite[Section 7.7.4]{nielsen2011quantum} 
as well as~\cite{cramer2010efficient,gross2010quantum,schmied2016qst}).

\subsection{Quantum gates}\label{subsec:basics_gates}

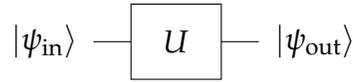
\begin{figure}
    \centerline{    
\newcommand\linelength{0.5}
\newcommand\verticaldiff{0.5}
\begin{tikzpicture}[scale = 1]

\node (lu) at (0,0) {};
\node (rl) at ($(lu)+(1.2,-1)$) {};
\draw ($(lu)$) rectangle ($(rl)$);
\node at ($(lu)+(0.6,-0.5)$) {\Large $U$};

\draw ($(lu)+(-\linelength,-\verticaldiff)$) -- ($(lu)+(0,-\verticaldiff)$);
\node at ($(lu)+(-\linelength-0.65,-\verticaldiff)$) {\Large $\ket{\psi_{\mathrm{in}}}$};
\draw ($(rl)+(\linelength,\verticaldiff)$) -- ($(rl)+(0,\verticaldiff)$);
\node at ($(rl)+(\linelength+0.75,\verticaldiff)$) {\Large $\ket{\psi_{\mathrm{out}}}$};
\end{tikzpicture}%
    }
    \caption{Circuit representation of a quantum gate acting on the input state $\ket{\psi_{\mathrm{in}}}$ and producing the output state 
    $\ket{\psi_{\mathrm{out}}}$. 
    Mathematically, the quantum gate is characterized via multiplication by the unitary matrix $U$, that is,
    $\ket{\psi_{\mathrm{out}}}=U\ket{\psi_{\mathrm{in}}}$.}
    \label{fig:quantum_gate}
\end{figure}

Classical computers consist of a collection of elementary logic gates (for example, AND, OR, NOT), which 
are applied to strings of classical bits.
Similarly, computations on quantum computers are represented by \emph{quantum gates},
which are applied to quantum states.
Mathematically, quantum gates are unitary matrices $U\in\bbu^{2^n}$, that is,
elements $U$ of $\bbc^{2^n\times 2^n}$ such that $U^\dagger U=I$, where $n$
is the number of qubits on which $U$ acts.
The action of the gate $U$ on the state $\ket{\psi_{\mathrm{in}}}$ is given by multiplication, that is, 
\begin{align}\label{eq:quantum_gate}
    \ket{\psi_{\mathrm{out}}}=U\ket{\psi_{\mathrm{in}}}.
\end{align}
Figure~\ref{fig:quantum_gate} provides a standard graphical illustration of the action of $U$ on $\ket{\psi_{\mathrm{in}}}$.
Applying the gate $U$ to the state $\ket{\psi_{\mathrm{in}}}$ constitutes a simple example of a quantum algorithm (equivalently referred to as quantum circuit), which we introduce in a more general form later in the paper.
We refer to the graphical scheme shown in Figure~\ref{fig:quantum_gate} as the circuit representation of the gate $U$.
By convention, quantum circuits are always read from left to right.
A quantum gate $U$ can be equivalently represented in terms of its Hermitian generator $H_{\rmU}=H_{\rmU}^\dagger$:
\begin{align}\label{eq:matrix_exponential}
    U=e^{-iH_{\rmU}}.
\end{align}
The matrix $H_{\rmU}$ is commonly referred to as \emph{Hamiltonian} due its 
physical meaning (compare ``Programming and experimental realization of quantum algorithms'').

    The unitarity of quantum gates has several interesting implications.
    First, note that quantum gates are \emph{linear} operators, which restricts the 
    range of possible computations on a quantum computer.
    Further, any quantum gate $U$ is reversible with the inverse being its Hermitian conjugate $U^\dagger$.  
    Note the fundamental difference to classical computing, which is \emph{not} reversible (consider the AND gate).
    Nevertheless, quantum computers can execute arbitrary classical algorithms based on the \emph{Toffoli} gate (the quantum analog of the NAND gate) and 
    auxiliary qubits (referred to as \emph{ancilla} qubits), see~\cite[Section 1.4.1]{nielsen2011quantum} for details.
    There are several further fascinating phenomena in quantum computing which are connected to the unitarity of quantum gates.
    One noteworthy example is the no-cloning theorem, which states 
    that it is not possible to copy qubits, that is, there exists no unitary matrix mapping 
    $\ket\psi\otimes\ket0$ to $\ket\psi\otimes\ket\psi$ for arbitrary $\ket{\psi}$, see~\cite[Box 12.1]{nielsen2011quantum} for details.
    Finally, note that the group of unitary matrices is a \emph{Lie group} with the associated \emph{Lie algebra} of skew-Hermitian matrices, compare~\eqref{eq:matrix_exponential}.
    This has useful implications in quantum control, see~\cite{dong2010quantum,altafini2012modeling,koch2022quantum} and the references therein for details.

    In ``Examples of Single-Qubit Gates'' and ``Examples of Multi-Qubit Gates'', we provide examples of commonly used quantum gates.

\begin{sidebar}{Examples of Single-Qubit Gates}

    \setcounter{sequation}{8}
    \renewcommand{\thesequation}{S\arabic{sequation}}
    \setcounter{stable}{0}
    \renewcommand{\thestable}{S\arabic{stable}}
    \setcounter{sfigure}{0}
    \renewcommand{\thesfigure}{S\arabic{sfigure}}

    \sdbarinitial{S}ingle-qubit gates are unitary $2\times2$-matrices.
The most common single-qubit gates are the \emph{Pauli gates} which are defined as
\begin{sequation}
X=\begin{bmatrix}
    0&1\\1&0
\end{bmatrix},\>
Y=\begin{bmatrix}
    0&-i\\i&0
\end{bmatrix},\>
Z=\begin{bmatrix}
    1&0\\0&-1
\end{bmatrix}.
\end{sequation}
The Pauli-$X$ gate is the quantum analog of the classical NOT gate since 
\begin{sequation}
X\ket0=\begin{bmatrix}0&1\\1&0\end{bmatrix}
\begin{bmatrix}
    1\\0
\end{bmatrix}
=
\begin{bmatrix}
    0\\1
\end{bmatrix}=
\ket1,
\end{sequation}
\begin{sequation}
X\ket1=\begin{bmatrix}
    0&1\\1&0
\end{bmatrix}
\begin{bmatrix}
    0\\1
\end{bmatrix}=
\begin{bmatrix}
    1\\0
\end{bmatrix}=\ket0.
\end{sequation}
The effect of single-qubit gates can be illustrated on the Bloch sphere.
For example, the Pauli-$X$ gate corresponds to a rotation of the input qubit around the $x$-axis with angle $\pi$ (and similarly for 
Pauli-$Y$ and Pauli-$Z$).
Rotations around different angles are possible as well:
The gates
\begin{sequation}\label{eq:single_qubits_rotations}
R_\rmx(\theta)=e^{-i\frac{\theta}{2}X},\>
R_\rmy(\theta)=e^{-i\frac{\theta}{2}Y},\>
R_\rmz(\theta)=e^{-i\frac{\theta}{2}Z}
\end{sequation}
rotate the input qubit around the $x$-, $y$-, and $z$-axis, respectively, by an angle of $\theta$.
Note that this is consistent with the above interpretation of the Pauli-$X$ gate since 
$R_\rmx(\pi)=e^{-i\frac{\pi}{2}X}=-iX$, that is, the two gates 
$R_\rmx(\pi)$ and $X$ are equivalent up to a global phase.

\sdbarfig{\includegraphics[width=0.7\columnwidth]{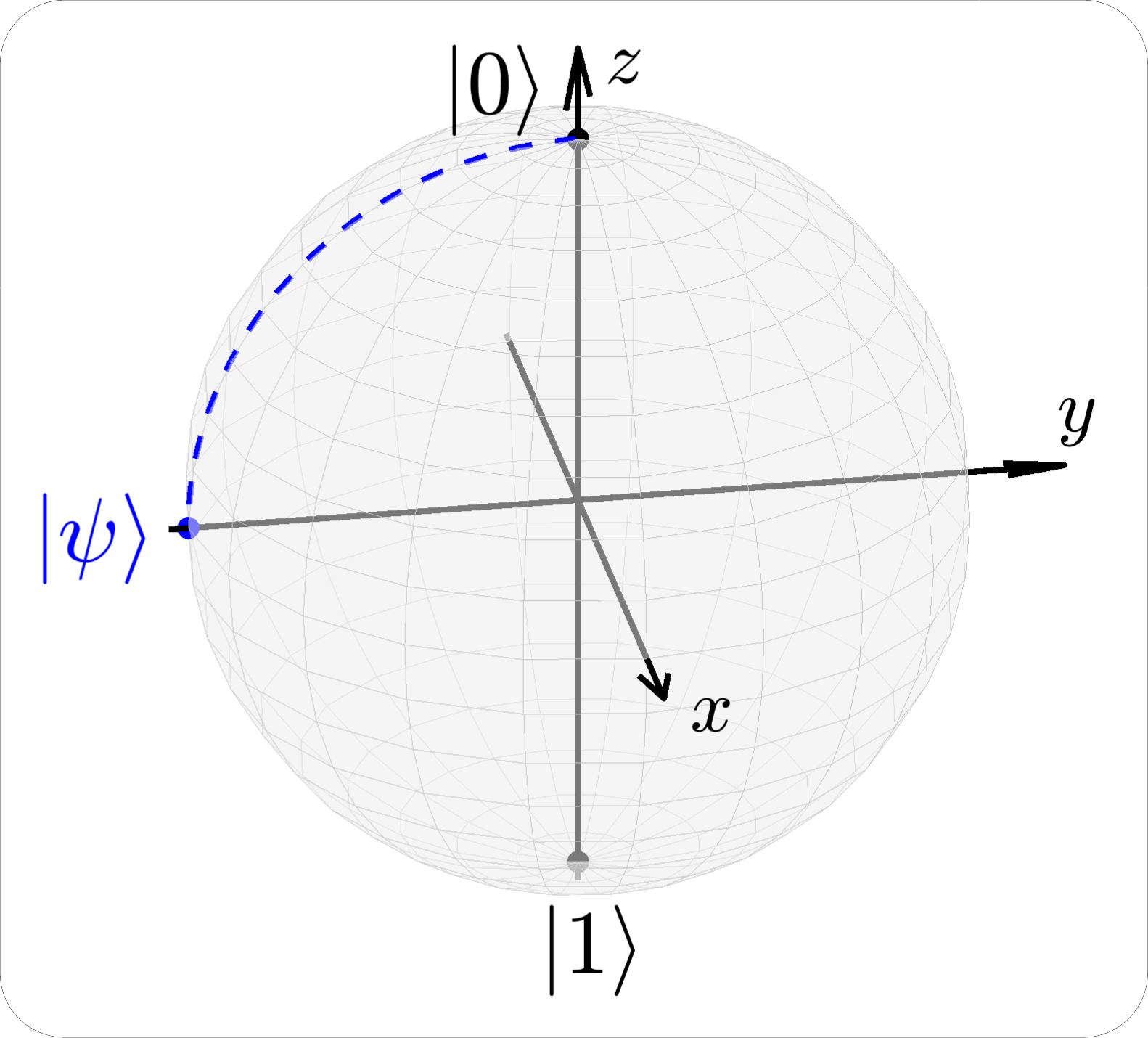}
}{Illustration of the rotation $R_{\rmx}(\frac{\pi}{2})$ applied to the qubit state $\ket0$ on the 
Bloch sphere, compare~\eqref{eq:bloch_sphere_angles}.
The final state is given by $\ket{\psi}=R_{\rmx}(\frac{\pi}{2})\ket0$.
The blue dashed curve depicts the qubit states $R_{\rmx}(\theta)\ket0$, where the parameter $\theta$ varies in $[0,\frac{\pi}{2}]$.\label{fig:bloch_sphere_Rx}}

In fact, arbitrary single-qubit gates can be interpreted as rotations on the Bloch sphere around \emph{some} axis.
Figure~\ref{fig:bloch_sphere_Rx} visualizes a rotation of the initial qubit in state $\ket0$ by the angle $\frac{\pi}{2}$ around the 
$x$-axis.
Another important gate is the \emph{Hadamard gate} defined as
\begin{sequation}\label{eq:Hadamard}
H=\frac{1}{\sqrt{2}}\begin{bmatrix}
        1&1\\1&-1
    \end{bmatrix}.
\end{sequation}
The main feature of the Hadamard gate is that it creates superposition, that is,
\begin{sequation}\label{eq:Hadamard_gate_plus}
H\ket0=\frac{1}{\sqrt{2}}(\ket0+\ket1)\eqqcolon\ket{+},
\end{sequation}
\begin{sequation}\label{eq:Hadamard_gate_minus}
H\ket1=\frac{1}{\sqrt{2}}(\ket0-\ket1)\eqqcolon\ket{-}.
\end{sequation}
When applying the Hadamard gate to one of the computational basis states $\ket0$ or $\ket1$, the result is $\ket{+}$ or $\ket{-}$, respectively.
These states both are an equal superposition of $\ket0$ and $\ket1$ and they form an alternative basis for single qubits.
The Hadamard gate can be visualized as a rotation by the angle $\pi$ around a tilted axis, compare Figure~\ref{fig:bloch_sphere_Hadamard}.

\sdbarfig{\includegraphics[width=0.7\columnwidth]{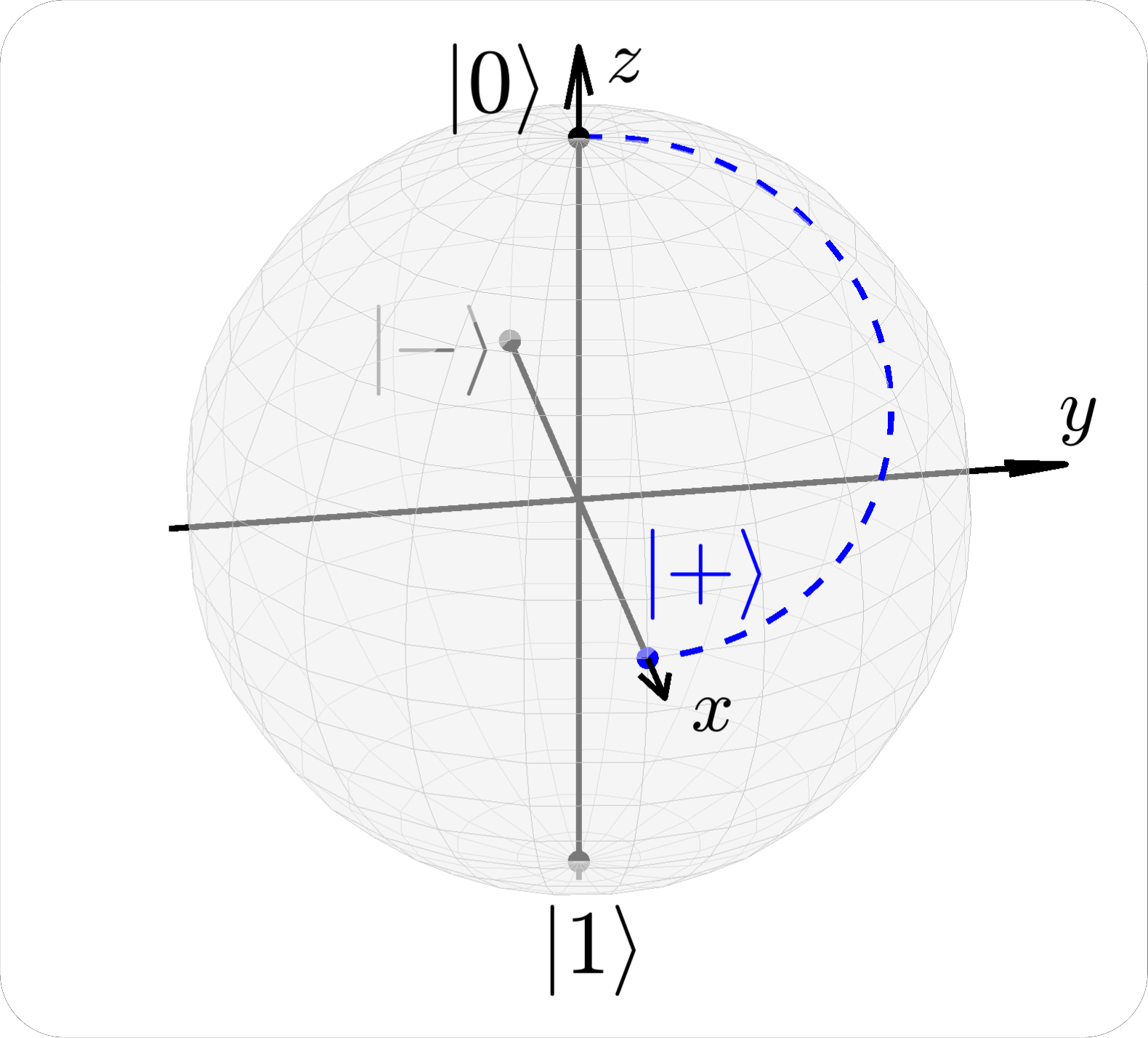}
}{Illustration of the Hadamard gate $H$ applied to the qubit state $\ket0$ on the 
Bloch sphere, compare~\eqref{eq:bloch_sphere_angles}, along with the two states $\ket{+}=\frac{1}{\sqrt{2}}(\ket0+\ket1)$ and $\ket{-}=\frac{1}{\sqrt{2}}(\ket0-\ket1)$.
The final state is given by $\ket{+}=H\ket0$.
The blue dashed curve depicts the qubit states $e^{-i\theta H_{\rmH}}\ket0$, where the parameter $\theta$ varies in $[0,\frac{\pi}{2}]$ and $H_{\rmH}$ is the Hermitian generator of $H$, that is, $H=e^{-iH_{\rmH}}$.\label{fig:bloch_sphere_Hadamard}}

\end{sidebar}

\begin{sidebar}{Examples of Multi-Qubit Gates}

\setcounter{sequation}{15}
\renewcommand{\thesequation}{S\arabic{sequation}}
\setcounter{stable}{0}
\renewcommand{\thestable}{S\arabic{stable}}
\setcounter{sfigure}{2}
\renewcommand{\thesfigure}{S\arabic{sfigure}}

\sdbarinitial{T}he design of meaningful quantum algorithms requires the use of quantum gates acting on multiple qubits.
First, we note that any single-qubit gate gives rise to a trivial multi-qubit gate, simply by leaving the other qubits unchanged.
Figure~\ref{fig:quantum_gate_multi_qubit} shows the circuit representation of applying the single-qubit gate $U$ to the second qubit
but leaving all other qubits unchanged.

\sdbarfig{
\newcommand\linelength{0.5}
\newcommand\verticaldiff{0.4}
\begin{tikzpicture}
\node (lu) at (0,0) {};
\node (rl) at ($(lu)+(1,-0.8)$) {};
\draw ($(lu)$) rectangle ($(rl)$);
\node at ($(lu)+(0.5,-0.4)$) {\Large $U$};
\draw ($(lu)+(-\linelength,-\verticaldiff)$) -- ($(lu)+(0,-\verticaldiff)$);
\node at ($(lu)+(-\linelength-0.75,-\verticaldiff-0.4)$) {\Large $\ket{\psi_{\mathrm{in}}}$};
\draw ($(rl)+(\linelength,\verticaldiff)$) -- ($(rl)+(0,\verticaldiff)$);
\node at ($(rl)+(\linelength+0.85,\verticaldiff-0.4)$) {\Large $\ket{\psi_{\mathrm{out}}}$};
\draw (-0.5,-1.2) -- (1.5,-1.2);
\draw (-0.5,0.4) -- (1.5,0.4);
\draw (-0.5,-2) -- (1.5,-2);
\end{tikzpicture}%
}{Circuit representation of the four-qubit gate $I\otimes U\otimes I\otimes I$ with a single-qubit gate $U$.
This quantum gate corresponds to applying $U$ to the second qubit and leaving the other three qubits unchanged.
For example, suppose the quantum state $\ket{\psi}$ is a product state, that is, $\ket{\psi}=\ket{\psi_1}\otimes\ket{\psi_2}\otimes\ket{\psi_3}\otimes\ket{\psi_4}$.
Then, applying the gate $I\otimes U\otimes I\otimes I$ to $\ket{\psi}$ produces the state 
$\ket{\psi_1}\otimes U\ket{\psi_2}\otimes\ket{\psi_3}\otimes\ket{\psi_4}$.\label{fig:quantum_gate_multi_qubit}}

Mathematically, the four-qubit gate describing this operation can be defined as 
\begin{sequation}
I_2\otimes U\otimes I_2\otimes I_2\in\bbu^{2^4}.
\end{sequation}
A frequently used alternative notation for only applying $U$ to the $j$-th qubit is 
\begin{sequation}\label{eq:U_j_notation_definition}
U_j\coloneqq \underbrace{I_2\otimes\dots\otimes I_2}_{j-1\>\text{times}}\otimes \> U\otimes 
\underbrace{I_2\otimes\dots\otimes I_2}_{n-j\>\text{times}}\in\bbu^{2^n}.
\end{sequation}

The most prominent non-trivial multi-qubit gate is the Controlled NOT (CNOT).
The CNOT is a two-qubit gate and its circuit representation is shown in Figure~\ref{fig:quantum_gate_CNOT}.

\sdbarfig{
\begin{tikzpicture}
\draw (0,0) -- (2,0);
\draw (0,1) -- (2,1);    
\draw (1,1) -- (1,-0.2);
\filldraw (1,1) circle (0.1cm);
\draw (1,0) circle (0.2cm);    
\node at (-0.65,0.5) {\Large $\ket{\psi_{\mathrm{in}}}$};
\node at (2.75,0.5) {\Large $\ket{\psi_{\mathrm{out}}}$};
\end{tikzpicture}%
}{Circuit representation of the CNOT gate.
The CNOT is the most prominent non-trivial multi-qubit gate.
It plays an important role in many quantum algorithms since it creates entanglement.
\label{fig:quantum_gate_CNOT}}

It is defined via the unitary matrix 
\begin{sequation}
\mathrm{CNOT}=\begin{bmatrix}
    1&0&0&0\\0&1&0&0\\0&0&0&1\\0&0&1&0
\end{bmatrix}.
\end{sequation}

The CNOT acts on the two qubits as follows:
It always leaves the first qubit (the upper qubit in Figure~\ref{fig:quantum_gate_CNOT}, called 
``control qubit'') unchanged.
Depending on the value of the control qubit, the second qubit (the lower qubit in Figure~\ref{fig:quantum_gate_CNOT}) is changed.
If the control qubit is $\ket0$, nothing happens, but if it is $\ket1$, a Pauli-$X$ (that is, a NOT) is applied to
the second qubit.
Indeed, it is simple to show that, for arbitrary $\ket{\psi}$,
\begin{sequation}
\mathrm{CNOT}(\ket0\otimes\ket{\psi})=\ket0\otimes\ket{\psi},
\end{sequation}
\begin{align*}
\mathrm{CNOT}(\ket1\otimes\ket{\psi})&=\ket1\otimes X\ket{\psi}.    
\end{align*}
The main feature of the CNOT gate is that it creates entanglement.
Suppose we are given the computational basis state $\ket0\otimes\ket0$ and we apply a Hadamard gate to the first qubit, 
leading to $\ket{+}\otimes\ket0$, compare~\eqref{eq:Hadamard_gate_plus}.
It is simple to show that
\begin{sequation}
\mathrm{CNOT}(\ket+\otimes\ket0)=\ket{\Phi^+}
\end{sequation}
with the Bell state $\ket{\Phi^+}=\frac{1}{\sqrt{2}}(\ket{00}+\ket{11})$, which was defined in~\eqref{eq:Bell_states} as a prime example 
of an entangled state.
Intuitively, applying a CNOT gate to a separable two-qubit state $\ket{\psi_1}\otimes\ket{\psi_2}$, where $\ket{\psi_1}$ is in superposition,
can create an entangled state.

For any single-qubit gate $U$, one can define a \emph{controlled-$U$ gate} (also denoted by $CU$):
The value of the control qubit (the first qubit) is not affected by $CU$.
If the control qubit is $\ket1$, then $U$ is applied to the second qubit, whereas, if the control qubit is $\ket0$, then the second qubit is left unchanged.
The circuit representation of a controlled-$U$ gate is shown in Figure~\ref{fig:quantum_gate_CU}.

\sdbarfig{
\begin{tikzpicture}
\draw (0,0) -- (0.5,0);
\draw (1.5,0) -- (2,0);
\draw (0,1.2) -- (2,1.2);    
\draw (1,1.2) -- (1,0.4);
\filldraw (1,1.2) circle (0.1cm);
\draw (0.5,0.4) rectangle (1.5,-0.4);
\node at (1,0) {\Large $U$};
\node at (-0.65,0.6) {\Large $\ket{\psi_{\mathrm{in}}}$};
\node at (2.75,0.6) {\Large $\ket{\psi_{\mathrm{out}}}$};
\end{tikzpicture}%
}{Circuit representation of a controlled-$U$ gate.
\label{fig:quantum_gate_CU}}

For a unitary matrix $U=\begin{bmatrix}
u_{11}&u_{12}\\u_{21}&u_{22}
\end{bmatrix}$, the matrix representation of the $CU$ gate is
\begin{sequation}\label{eq:controlled_U}
CU=\begin{bmatrix}
    1&0&0&0\\0&1&0&0\\0&0&u_{11}&u_{12}\\0&0&u_{21}&u_{22}
\end{bmatrix}.
\end{sequation}

Finally, we introduce the SWAP gate which is another important $2$-qubit gate.
Its circuit representation is shown in Figure~\ref{fig:quantum_gate_SWAP}.

\sdbarfig{
\begin{tikzpicture}
\draw (0,0) -- (2,0);
\draw (0,1) -- (2,1);    
\draw (1,1) -- (1,0); 
\draw (0.8,1.2)--(1.2,0.8);
\draw (0.8,0.8)--(1.2,1.2);
\draw (0.8,0.2)--(1.2,-0.2);
\draw (0.8,-0.2)--(1.2,0.2);
\node at (-0.65,0.5) {\Large $\ket{\psi_{\mathrm{in}}}$};
\node at (2.75,0.5) {\Large $\ket{\psi_{\mathrm{out}}}$};
\end{tikzpicture}%
}{Circuit representation of the SWAP gate.
\label{fig:quantum_gate_SWAP}}

The corresponding unitary matrix is
\begin{sequation}
\mathrm{SWAP}=\begin{bmatrix}
    1&0&0&0\\0&0&1&0\\0&1&0&0\\0&0&0&1
\end{bmatrix}.
\end{sequation}
If the input state is separable, that is, of the form $\ket{\psi_1}\otimes\ket{\psi_2}$ for two single-qubit states $\ket{\psi_1}$ and $\ket{\psi_2}$, then the SWAP gate, quite literally, swaps the two states and produces
\begin{sequation}\label{eq:SWAP}
\mathrm{SWAP}(\ket{\psi_1}\otimes\ket{\psi_2})=\ket{\psi_2}\otimes\ket{\psi_1}.
\end{sequation}

\end{sidebar}

    \begin{figure}
        \centerline{
            \newcommand\pqcdiff{0.5}
            \newcommand\pqcleft{0.4}
            \begin{tikzpicture}[scale = 1.2]
            \node (lu) at (0,3) {};
            \node (ru) at ($(lu)+(2,0)$) {};
            \draw ($(lu)$) rectangle ($(lu)+(2,-3)$);
            \node at ($(lu)+(1,-1.5)$) {\huge $U$};
            
            \draw ($(lu)+(-\pqcleft,-\pqcdiff)$) -- ($(lu)+(0,-\pqcdiff)$);
            \draw ($(lu)+(-\pqcleft,-\pqcdiff-\pqcdiff)$) -- ($(lu)+(0,-\pqcdiff-\pqcdiff)$);
            \node at ($(lu)+(-\pqcleft+0.175,-\pqcdiff-\pqcdiff-\pqcdiff-0.28*\pqcdiff)$) {\large$\vdots$};
            \draw ($(0,0)+(-\pqcleft,\pqcdiff)$) -- ($(0,0)+(0,\pqcdiff)$);
            \node at ($(ru)+(-3,-1.5)$) {\Large $\ket{\psi_0}$};
            
            \draw ($(ru)+(0,-\pqcdiff)$) -- ($(ru)+(\pqcleft,-\pqcdiff)$);
            \node at ($(ru)+(\pqcleft+0.26,-\pqcdiff)$) {\includegraphics[width=0.25in]{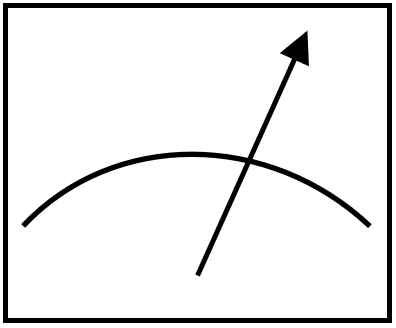}};
            \draw ($(ru)+(0,-\pqcdiff-\pqcdiff)$) -- ($(ru)+(\pqcleft,-\pqcdiff-\pqcdiff)$);
            \node at ($(ru)+(\pqcleft+0.26,-\pqcdiff-\pqcdiff)$) {\includegraphics[width=0.25in]{Figures/measurement}};
            \draw ($(2,0)+(0,\pqcdiff)$) -- ($(2,0)+(\pqcleft,\pqcdiff)$);
            \node at ($(2,0)+(\pqcleft+0.26,\pqcdiff)$) {\includegraphics[width=0.25in]{Figures/measurement}};
            \node at ($(ru)+(\pqcleft+0.26,-\pqcdiff-\pqcdiff-\pqcdiff-0.28*\pqcdiff)$) {$\vdots$};
            
            \node at ($(ru)+(1.5,-1.5)$) {\Large $\calM$};
            \end{tikzpicture}%
        }
        \caption{A generic quantum algorithm including an input state $\ket{\psi_0}$, a unitary matrix $U$, and 
        a projective measurement with respect to the observable $\calM$.
        The unitary matrix $U$ usually consists of parallel and series interconnections of smaller quantum gates (typically single- or two-qubit gates).    
        }
        \label{fig:quantum_algorithm}
    \end{figure}

\begin{sidebar}{Programming and experimental realization of quantum algorithms}

    \setcounter{sequation}{23}
    \renewcommand{\thesequation}{S\arabic{sequation}}
    \setcounter{stable}{0}
    \renewcommand{\thestable}{S\arabic{stable}}
    \setcounter{sfigure}{6}
    \renewcommand{\thesfigure}{S\arabic{sfigure}}

    \sdbarinitial{I}n the main text, we have introduced the basic framework of quantum computing on an abstract mathematical level.
    In the following, we discuss how quantum algorithms are implemented and executed on real quantum hardware.
    We address key issues connected to both the algorithmic implementation and the 
experimental realization of quantum computing.

Let us start by explaining why quantum gates are unitary.
The time evolution of a quantum state $\ket{\psi(t)}$ is described by the 
famous \emph{Schr\"odinger equation}~\cite[Postulate 2']{nielsen2011quantum}
\begin{sequation}\label{eq:schroedinger_equation}
    \frac{\rmd}{\rmd t}\ket{\psi(t)}=-iH\ket{\psi(t)}
\end{sequation}
for $t\geq0$.
Here, $H$ is the
\emph{Hamiltonian} which describes the given physical system.
Despite the notational overlap,
the Hamiltonian $H$ of a quantum system is not to be confused with the 
Hadamard gate $H$ which is defined as the unitary matrix in~\eqref{eq:Hadamard}.
The solution $\ket{\psi(t)}$, $t\geq0$, of the Schr\"odinger equation from an initial condition $\ket{\psi_0}$ is 
\begin{sequation}
    \ket{\psi(t)}=e^{-iHt}\ket{\psi_0},\>t\geq0.
\end{sequation}
Recall that $e^{-iHt}$ is unitary and, conversely, for any unitary matrix $U$, there exist
$H=H^\dagger$, $t\geq0$ satisfying $U=e^{-iHt}$.
Hence, a unitary $U$ can be applied to the quantum state $\ket{\psi}$ by letting some quantum system evolve from the initial condition $\ket{\psi}$ under a specific Hamiltonian $H$ for some time $t\geq0$ (that is, such that $U=e^{-iHt}$).

Building a quantum computer to realize this unitary evolution on real quantum hardware is a challenging task, and different approaches have been developed.
In the following, we mention three possibilities and we refer to~\cite[Section 7]{nielsen2011quantum}
for details and further alternatives.
Superconducting qubits~\cite{kjaergaard2020superconducting_sidebar} are a promising candidate and were used in a number of recent milestone
experiments, for example,~\cite{arute2019quantum,kim2023evidence}.
Here, the qubit is implemented as an LC circuit with superconducting material.
The circuit is cooled down close to absolute zero in order to avoid dissipation losses and enable the occurrence of quantum effects.
Quantum gates can then be implemented via microwave pulses with specific phases and frequencies.
Another popular approach are trapped-ion quantum computers~\cite{bruzewicz2019trapped_sidebar}.
In these devices, the qubit is encoded into the states of a charged particle which can be influenced 
via lasers to realize quantum gates.
Alternatively, photonic quantum computing~\cite{slussarenko2019photonic_sidebar} relies on photons to represent qubits.
Photons have a number of desirable properties such as simplicity of transmitting and influencing them on the single-qubit level,
but implementing, for example, two-qubit gates is non-trivial.

In each of these approaches, it is not possible to directly realize an arbitrary unitary matrix
on the hardware level.
Instead, experimental realizations provide a limited set of gates.
Typically, this \emph{basic gate set} (or \emph{elementary gate set}) consists of a few single-qubit operations, 
for example, Pauli rotations, and at least one multi-qubit gate, for example, CNOT.
For suitable choices of the basic gate set, one can show that parallel and series interconnections of the gates 
allow to approximate a given quantum algorithm, that is, a unitary matrix, up to an arbitrary precision.
Such gate sets are called \emph{universal}, compare\ \cite[Section 4.5]{nielsen2011quantum} for details.
One example is the gate set $\sqrt{X}$, $X$, CNOT, and $R_{\rmz}(\theta)$ with free parameter $\theta$, which is commonly 
used by IBM~\cite{ibm_gateset_sidebar}.

Thus, when implementing a quantum algorithm in practice, one first needs to approximate the unitaries used in the algorithm design
based on the available basic gate set.
This approximation is carried out during \emph{quantum circuit compilation}, which includes further steps to match the algorithm
to the available hardware (for example, taking the connectivity of the qubits into account)~\cite{larose2019overview_sidebar}.
At this stage, it is also possible to perform additional optimization in order to reduce the circuit depth 
(maximum number of sequentially applied gates) or width (number of qubits)~\cite{maslov2008quantum_sidebar}.
There exist different software frameworks for performing these steps:
Quantum algorithms can be programmed using familiar programming languages, 
there exist automatic compilers, and they can even be executed on 
actual quantum computers via the cloud~\cite{leymann2020quantum_sidebar}.
We refer to~\cite{larose2019overview_sidebar} for an overview on these points and further practical issues.   
Figure~\ref{fig:realization_scheme} provides a high-level summary of the steps required for implementing a quantum algorithm.

\sdbarfig{\includegraphics[width=0.95\columnwidth]{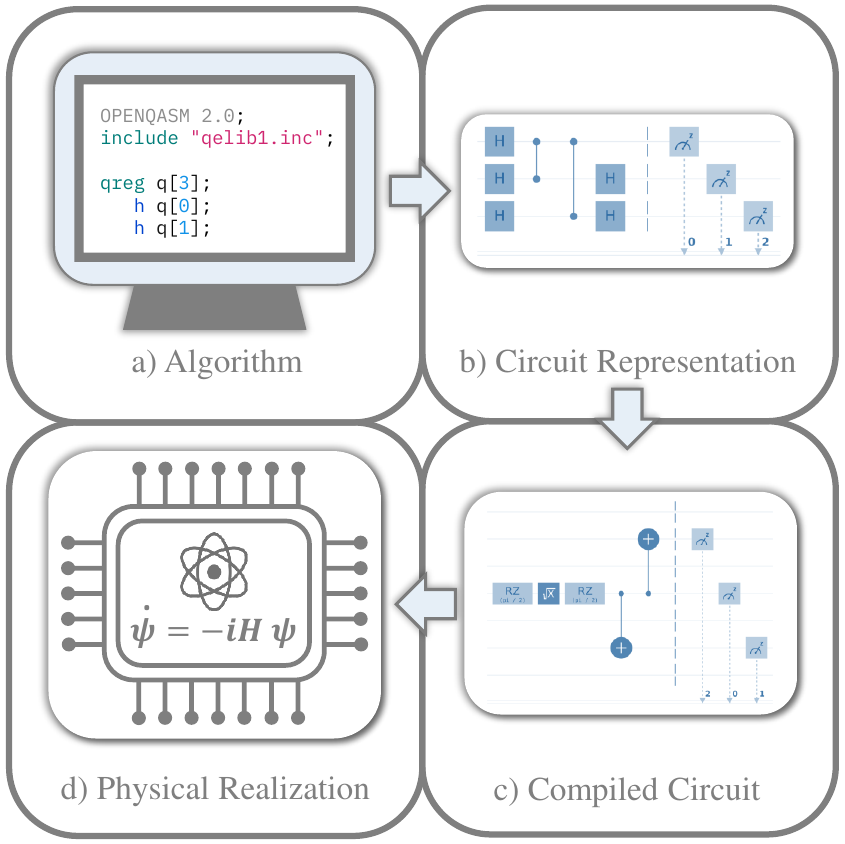}
}{Basic scheme summarizing the implementation of a quantum algorithm.
a) The first step is to program the quantum algorithm, for which different software frameworks exist.
b) The circuit representation is an equivalent graphical reformulation of the algorithm which can often be more insightful.
c) During compilation, the originally programmed circuit is transformed into a different circuit depending on, for example, the available basic gate set, the connectivity of the qubits, and other hardware requirements.
d) Finally, the compiled circuit is executed in the real world via the time evolution of a quantum system.
\label{fig:realization_scheme}}
\end{sidebar}

\begin{sidebar}{\continuesidebar}
    \setcounter{sequation}{25}
    \renewcommand{\thesequation}{S\arabic{sequation}}
    \setcounter{stable}{0}
    \renewcommand{\thestable}{S\arabic{stable}}
    \setcounter{sfigure}{8}
    \renewcommand{\thesfigure}{S\arabic{sfigure}}

    \end{sidebar}

    \section{Quantum algorithms}\label{sec:quantum_algorithm}

    In the previous section, we have defined all the basic ingredients of a quantum computer.
    Now, we combine these ingredients to form a \emph{quantum algorithm} (also referred to as \emph{quantum circuit}).
    Quantum algorithms are computational schemes that can be executed on quantum computers.
    Examples include Grover's search algorithm~\cite{grover1996fast} and Shor's algorithm for integer factorization~\cite{shor1999polynomial}.
    In what follows, we focus on the key mathematical definition and properties of quantum algorithms.
    In ``Programming and experimental realization of quantum algorithms'', we discuss how quantum algorithms can be implemented and executed on real quantum hardware.

A quantum algorithm consists of three main building blocks:
an input state, quantum gates, and measurement.
Figure~\ref{fig:quantum_algorithm} shows a generic quantum algorithm consisting of these three building blocks.
The input state is denoted by $\ket{\psi_0}$, and in many cases it is chosen as the computational basis state 
\begin{align}
    \ket{0}^{\otimes n}=\underbrace{\ket0\otimes\dots\otimes\ket0}_{n\>\text{times}}.
\end{align}
Mathematically, choosing $\ket0^{\otimes n}$ as input state is without loss of generality since a different input state $\ket{\psi_0}$
can always be rewritten as $\ket{\psi_0}=U'\ket0^{\otimes n}$ for some unitary $U'$ which can be included into the main unitary matrix $U$
of the algorithm.
From a practical perspective, choosing $\ket0^{\otimes n}$ is often a meaningful choice as well since it can often 
be generated more easily on physical devices.
The process of generating an input state to be used in a quantum algorithm is called \emph{state preparation}, and it essentially requires solving a \emph{quantum optimal control} problem, that is, an optimal control problem for a quantum mechanical system~\cite{koch2022quantum}.

Second, the algorithm shown in Figure~\ref{fig:quantum_algorithm} contains the unitary matrix $U$ which acts on the input state.
Typically, $U$ is a sequence of many smaller quantum gates, for example, of Pauli, Hadamard, or CNOT gates.
Indeed, note that parallel and series interconnections of quantum gates form again quantum gates.
More precisely, concerning series interconnections, the product $U_1U_2$ of two unitary matrices $U_1,U_2\in\bbu^{2^n}$ is again unitary since 
\begin{align}
    (U_1U_2)^\dagger U_1U_2=U_2^\dagger U_1^\dagger U_1U_2=I_{2^n}. 
\end{align}
Further, concerning parallel interconnections, the tensor product $U_1\otimes U_2$ of $U_1,U_2\in\bbu^{2^n}$ is also 
unitary since 
\begin{align}
    &(U_1\otimes U_2)^\dagger(U_1\otimes U_2)=
    (U_1^\dagger\otimes U_2^\dagger) (U_1\otimes U_2)\\\nonumber
    =&(U_1^\dagger U_1)\otimes(U_2^\dagger U_2)=I_{2^n}\otimes I_{2^n}=I_{2^{2n}}.
\end{align}
Unfortunately, there does not seem to be a simple possibility to interconnect quantum gates via feedback~\cite[p. 23]{nielsen2011quantum}.

\begin{figure}
    \centerline{
        \begin{tikzpicture}[scale = 1]
        \node at (-0.6,0) {\large$\ket{\psi_0}$};
        \draw (0,0)--(1,0);
        \node at (1.3,0) {\includegraphics[width=0.25in]{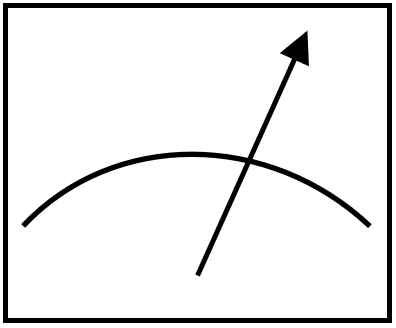}};
        \end{tikzpicture}%
    }
    \caption{Circuit representation of a projective measurement 
    in the computational basis.
    Performing this measurement} amounts to determining, for each qubit, whether it is in state $\ket0$ or $\ket1$.
    \label{fig:quantum_algorithm_measurement}
\end{figure}

Finally, to access the result of a quantum algorithm, measurements are required.
In quantum algorithms, measurements are represented as in Figure~\ref{fig:quantum_algorithm_measurement},
which shows the trivial quantum algorithm consisting only of a measurement of the input state $\ket{\psi_0}$.
The algorithm shown in Figure~\ref{fig:quantum_algorithm} performs a projective measurement with observable $\calM$ 
at the end.
If no observable is given in the circuit (as in Figure~\ref{fig:quantum_algorithm_measurement}), then measurements are typically understood in the computational basis, that is, each
individual qubit is measured with respect to the observable Z.

It turns out that, in many cases, there is a simple mathematical formula describing the full quantum algorithm.
Recall that projective measurements can be understood as (statistical estimates of) the quadratic form $\braket{\psi|\calM|\psi}$, compare~\eqref{eq:measurement_quadratic_function}.
In many quantum algorithms (especially in VQAs, see 
the section ``Variational quantum algorithms''), this quadratic form is the main quantity of interest and forms the output of the algorithm.
In this case, the overall algorithm can be summarized as
\begin{align}\label{eq:quantum_algorithm_quadratic_form}
    \braket{\psi_0|U^\dagger\calM U|\psi_0}.
\end{align}
Thus, despite the seemingly complicated nature of quantum mechanics, a large class of 
quantum algorithms can be summarized as a quadratic form acting on a unitary matrix acting on an input state.
The main non-trivial challenges are how $U$ and $\calM$ are chosen and how the algorithm leading to~\eqref{eq:quantum_algorithm_quadratic_form}
is implemented experimentally.
In general, the expression~\eqref{eq:quantum_algorithm_quadratic_form} cannot be efficiently simulated on a classical computer for 
relevant numbers of qubits since the size of the involved matrices is $\bbc^{2^n\times 2^n}$.
It is important to note, however, that this size only becomes problematic when the algorithm includes entanglement, that is, when at least some of the qubits are entangled.
Indeed, consider the extreme case where the input state is the tensor product of $n$ single qubits (for example, a computational basis state as in~\eqref{eq:composite_state_computational_basis}) and $U$ only consists of parallel and series interconnections of single-qubit gates.
In that case, the quantum algorithm can be easily simulated on a classical computer by considering each qubit separately.
To summarize, the main power of quantum algorithms as in~\eqref{eq:quantum_algorithm_quadratic_form} lies in a combination of the sheer size
of the involved quantities, which
grows exponentially with the number of qubits, together with entanglement.

We have introduced quantum algorithms on a generic, mathematical level.
At this point, it might be unclear how to work with quantum algorithms in practice, for example, how to design an algorithm that solves a meaningful problem.
Generally speaking, designing quantum algorithms is an art.
In control language, it is like finding a Lyapunov function for stability analysis:
If an algorithm is available, then verifying that it works is straightforward, but
finding an algorithm that solves a given problem in the first place is hard and there exists no general recipe.
Nevertheless, there are lots of insights into algorithm design and numerous powerful algorithms have been 
proposed in the literature.
The goal of this tutorial is to equip the reader with the mathematical basics for understanding quantum algorithms, 
but providing an in-depth introduction into algorithm design goes beyond our scope.
The interested reader is referred to~\cite{nielsen2011quantum} for a more detailed introduction to the topic.
In ``The Quantum Fourier Transform'', we discuss one of the most popular quantum algorithms.
Further, in ``Using Quantum Computers in Control'', we discuss possible applications of quantum computing by listing several algorithms that solve computational problems relevant in control.

\begin{sidebar}{The Quantum Fourier Transform}

    \setcounter{sequation}{25}
    \renewcommand{\thesequation}{S\arabic{sequation}}
    \setcounter{stable}{0}
    \renewcommand{\thestable}{S\arabic{stable}}
    \setcounter{sfigure}{7}
    \renewcommand{\thesfigure}{S\arabic{sfigure}}

    \sdbarinitial{W}e present the basic problem setup and the $3$-qubit circuit of the 
    quantum Fourier transform.
    Further details can be found in~\cite[Section 5]{nielsen2011quantum}.
    Recall that the classical discrete Fourier transform computes, for a sequence of complex numbers $\{x_k\}_{k=1}^N$,
    an output sequence $\{y_k\}_{k=1}^N$ via 
    \begin{sequation}
        y_k=\frac{1}{\sqrt{N}}\sum_{j=1}^N x_j e^{2\pi i\frac{jk}{N}}.
    \end{sequation}
    The quantum Fourier transform performs the same computation on quantum states.
    In the following, we use the common decimal notation of the computational basis states~\eqref{eq:composite_state_computational_basis}, meaning that, for example,
    \begin{sequation}
        \ket{9}=\ket{1001}=\ket1\otimes\ket0\otimes\ket0\otimes\ket1.
    \end{sequation}
    For the computational basis states $\ket0$, $\ket1$, $\dots$, $\ket{N-1}$, the quantum Fourier transform is defined via
    \begin{sequation}
        \ket{j}\mapsto\frac{1}{\sqrt{N}}\sum_{k=0}^{N-1}e^{2\pi i\frac{jk}{N}}\ket{k}.
    \end{sequation}
    The definition is extended to arbitrary quantum states by imposing linearity, leading to a map 
    \begin{sequation}
        \sum_{j=0}^{N-1}x_j\ket{j}\mapsto\sum_{k=0}^{N-1}y_k\ket{k}.
    \end{sequation}
    The amplitudes $y_k$ of the transformed state are equal to the classical discrete Fourier transform of the amplitudes $x_j$
    of the original state.
    It turns out that this operation is a unitary transformation and, therefore, it can be implemented on a quantum computer.
    The circuit for $3$ qubits is shown in Figure~\ref{fig:quantum_algorithm_QFT}. 
    In addition to the Hadamard gate $H$ from~\eqref{eq:Hadamard}, the circuit contains a SWAP gate (compare\ \eqref{eq:SWAP}) as well as
    controlled $S$ and $T$ gates (compare\ \eqref{eq:controlled_U}) with the phase gate 
    $S=\begin{bmatrix}
        1&0\\0&i
    \end{bmatrix}
$ and the $T$ gate $T=\begin{bmatrix}
    1&0\\0&\exp\left(i\frac{\pi}{4}\right)
\end{bmatrix} $.

\sdbarfig{
    \newcommand\linesep{1}
        \begin{tikzpicture}[scale=0.9]
            \draw (0,0)--(0.4,0);
            \draw (0.4,0.4) rectangle (1.2,-0.4);
            \node at (0.8,0) {\Large $H$};
            \draw (1.2,0)--(1.4,0);
            \draw (1.4,0.4) rectangle (2.2,-0.4);
            \node at (1.8,0) {\Large $S$};
            \draw (2.2,0)--(2.4,0);
            \draw (2.4,0.4) rectangle (3.2,-0.4);
            \node at (2.8,0) {\Large $T$};
            \draw (3.2,0)--(7,0);
            \draw (6.45,0.15)--(6.75,-0.15);
            \draw (6.45,-0.15)--(6.75,0.15);
            \draw (6.6,0)--(6.6,-2);
            \draw (0,-1)--(3.4,-1);
            \draw (3.4,-0.6) rectangle (4.2,-1.4);
            \node at (3.8,-1) {\Large $H$};
            \draw (4.2,-1)--(4.4,-1);
            \draw (4.4,-0.6) rectangle (5.2,-1.4);
            \node at (4.8,-1) {\Large $S$};
            \draw (5.2,-1)--(7,-1);
            \filldraw (1.8,-1) circle (0.1cm);
            \draw (1.8,-1)--(1.8,-0.4);
            \draw (0,-2)--(5.4,-2);
            \draw (5.4,-1.6) rectangle (6.2,-2.4);
            \node at (5.8,-2) {\Large $H$};
            \draw (6.2,-2)--(7,-2);
            \draw (6.45,-1.85)--(6.75,-2.15);
            \draw (6.45,-2.15)--(6.75,-1.85);
            \filldraw (2.8,-2) circle (0.1cm);
            \filldraw (4.8,-2) circle (0.1cm);
            \draw (2.8,-2)--(2.8,-0.4);
            \draw (4.8,-2)--(4.8,-1.4);
        \end{tikzpicture}%
}{Circuit of the $3$-qubit quantum Fourier transform.
The quantum Fourier transform performs a standard Fourier transform on the probability amplitudes of the input state.
It plays a crucial role as a subroutine in numerous quantum algorithms, including Shor's algorithm for integer factorization~\cite{shor1999polynomial}.\label{fig:quantum_algorithm_QFT}}

Given the wide range of applications of the Fourier transform, its implementation on a quantum computer 
promises far-reaching advancements.
It can be shown that, in order to transform a sequence of length $2^n$, the quantum Fourier transform 
only uses on the order of $\mathcal{O}(n^2)$ gates, whereas the classical Fast Fourier Transform requires $\mathcal{O}(n2^n)$ classical operations.
In practice, this exponential speedup cannot be easily exploited, however, since the output of the Fourier transform is encoded in the 
probability amplitudes of a quantum state, which are not directly accessible.
Nevertheless, the quantum Fourier transform is an important subroutine of several quantum algorithms and, for example, forms 
the basis for Shor's algorithm~\cite{shor1999polynomial}.
    \end{sidebar}

    \begin{sidebar}{Using Quantum Computers in Control}

        \setcounter{sequation}{29}
        \renewcommand{\thesequation}{S\arabic{sequation}}
        \setcounter{stable}{0}
        \renewcommand{\thestable}{S\arabic{stable}}
        \setcounter{sfigure}{0}
        \renewcommand{\thesfigure}{S\arabic{sfigure}}
        
        \sdbarinitial{T}his tutorial introduces quantum computing from the control perspective.
        The main motivation is that many research challenges in quantum computing are closely connected to control and, thus, control may contribute to advance the field of quantum computing (see ``Research Challenges in Quantum Computing'').
        In the following, we address the converse direction, that is, what quantum computing can contribute to control.
        To this end, we provide a short list of computational problems which are relevant in control and for which quantum algorithms have been developed.

        \subsection{Combinatorial optimization}
        Various quantum algorithms were developed to solve combinatorial optimization problems~\cite{sanders2020compilation_sidebar,gemeinhardt2023quantum_sidebar}.
        This includes the quantum approximate optimization algorithm (QAOA)~\cite{farhi2014quantum} explained in the main text as well as quantum annealing~\cite{hauke2020perspectives_sidebar}, which can be viewed as a continuous version of the discrete, gate-based quantum computing framework considered in this tutorial.
        Although there are only few theoretical results on the performance of these algorithms, and proving improvements over classical algorithms is an open problem, combinatorial optimization still provides one of the most promising use-cases of quantum computing in the near future.
    
        Optimization over integer variables is important in many branches of control.
        This includes problems with discrete actuators or discrete state-space appearing, for example, in hybrid systems~\cite{goebel2009hybrid_sidebar}, but also classical problems in systems theory and control such as the stability of interval matrices or static output-feedback~\cite{blondel2000survey}.
        All of these problems are hard to solve on a classical computer and, therefore, quantum computing has the potential to provide computational improvements.
        In the existing literature, only few contributions have investigated the use of quantum computing for solving combinatorial problems arising in control~\cite{inoue2020model_sidebar,deshpande2022quantum_sidebar,schneider2024using_sidebar}, which leaves numerous possibilities for future research.

        \subsection{Mixed-integer optimization}
        Many control applications, for example, in energy systems, mobility, or medicine admit both discrete and continuous decision variables, thus leading to mixed-integer optimization problems.
         Solving such problems on a quantum computer has been addressed recently in the literature, for example, by~\cite{gambella2020multiblock_sidebar,brown2022copositive_sidebar,chang2022hybrid_sidebar}.
         These approaches mostly rely on extensions of existing quantum algorithms for combinatorial optimization and, thus, they share their benefits and limitations.

        \subsection{Semidefinite programming}
        Semidefinite programming is a cornerstone of modern control theory~\cite{boyd1994linear_sidebar,scherer2000linear_sidebar}.
        However, solving semidefinite programs can be challenging for large-scale problems, for example, resulting from sum-of-squares optimization~\cite{ebenbauer2006analysis_sidebar}.
        In recent years, various quantum algorithms for semidefinite programming have been developed~\cite{brandao2016quantum_sidebar,kerenidis2020quantum_sidebar,patel2021variational_sidebar,augustino2023quantum_sidebar}, which may possibly be used to solve control problems more efficiently than previously possible.

        \subsection{Linear systems of equations}
        The Harrow-Hassidim-Lloyd (HHL) algorithm was developed by~\cite{harrow2009quantum_sidebar} for solving linear systems of equations.
        Under suitable assumptions, it admits a provable exponential speedup over its best known classical counterpart.
        Given that linear algebra is a key language for various domains of control theory, the HHL algorithm may be useful, especially for large-scale problems where its computational speedup becomes significant.

        \end{sidebar}

\section{Variational quantum algorithms}\label{sec:vqa}

\begin{pullquote}
    Variational quantum algorithms are feedback interconnections, consisting of a static nonlinearity and a discrete-time dynamical system.
\end{pullquote}

VQAs are a class of quantum algorithms which 
contain a parameterized quantum circuit, where the parameters are adapted iteratively via a classical optimization algorithm~\cite{cerezo2021variational}.
Mathematically, VQAs are feedback interconnections consisting of a discrete-time dynamical system with a static nonlinear function.
The motivation for VQAs stems from the fact that, in the current NISQ era, quantum devices are mostly small-scale and noisy, which poses
 severe challenges to the implementation of non-trivial quantum algorithms.
VQAs promise to perform meaningful computations already with few gates and qubits since the gates are optimized, 
which not only allows for a reduction of the overall size of the circuit but also for an adaptation to noise on the device.

VQAs are an active research field with numerous interesting and challenging problems, both from the theoretical and the practical side.
We refer to~\cite{cerezo2021variational} for a comprehensive survey.
In this section, we provide an introduction to VQAs.
After explaining the main concept,
we present three important examples of VQAs:
the \emph{variational quantum eigensolver} (VQE),
the \emph{quantum approximate optimization algorithm} (QAOA),
and \emph{quantum machine learning} (QML) based on VQAs.

\subsection{The main idea}\label{subsec:vqa_main_idea}

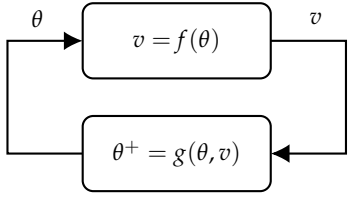
\begin{figure}
    \centerline{
        \begin{tikzpicture}[scale=1]
        \draw[-{Latex[length=2.5mm,width=2.5mm]},thick] (2,-1.5)--(1,-1.5)--(1,0)--(2,0);
        \draw[thick,rounded corners] (2,0.5) rectangle (4.5,-0.5);
        \node at (3.25,0) {$v=f(\theta)$};
        \draw[-{Latex[length=2.5mm,width=2.5mm]},thick] (4.5,0)--(5.5,0)--(5.5,-1.5)--(4.5,-1.5);
        \draw[thick,rounded corners] (2,-1) rectangle (4.5,-2);
        \node at (3.25,-1.5) {$\theta^+=g(\theta,v)$};
        \node at (1.4,0.3) {$\theta$};
        \node at (5.1,0.3) {$v$};
        \end{tikzpicture}%
    }
    \caption{Basic scheme of a variational quantum algorithm (VQA).  
VQAs are a feedback interconnection of a parameterized quantum circuit $f(\theta)$ with an update rule $\theta^+=g(\theta,v)$ for the parameter vector $\theta$.
Evaluating $f(\theta)$ requires executing the quantum algorithm with the unitary matrix $U(\theta)$ from~\eqref{eq:vqa_U_def} and the observable $\calM$, compare~\eqref{eq:vqa_f_definition}.
On the contrary, the update rule for $\theta$ is commonly implemented on a classical computer.
In many VQAs, $f(\theta)$ plays the role of a cost function, in which case the update rule $g(\theta,v)$ is chosen such that $\theta$ iteratively converges to a (local) minimum of $f$.
}
    \label{fig:vqa}
\end{figure}

VQAs involve a \emph{parameterized quantum circuit}, that is, 
a parameterized unitary matrix $U(\theta)$ of the form
\begin{align}\label{eq:vqa_U_def}
    U(\theta)=U_1(\theta_1)U_2(\theta_2)\cdots U_N(\theta_N)
\end{align}
with individual parameterized unitaries $U_j:\bbr\to\bbu^{2^n}$, $j=1,\dots,N$.
Each $U_j(\theta_j)$, $j=1,\dots,N$, takes the form
\begin{align}\label{eq:vqa_Ui_def}
    U_j(\theta_j)=e^{-i\theta_jH_j}
\end{align}
for some Hermitian generator $H_j=H_j^\dagger\in\bbc^{2^n\times 2^n}$.
For a fixed parameter $\theta\in\bbr^N$, we apply $U(\theta)$ to an input state $\ket{\psi_0}$ and, subsequently, 
evaluate the quadratic form corresponding to the projective measurement with observable $\calM$, compare~\eqref{eq:quantum_algorithm_quadratic_form}.
This quantum algorithm gives rise to a static nonlinear function $f(\theta)$ which, for $\theta\in\bbr^N$, returns the value
\begin{align}\label{eq:vqa_f_definition}
    f(\theta)=\braket{\psi_0|U(\theta)^\dagger\calM U(\theta)|\psi_0}.
\end{align}
VQAs are feedback interconnections of $f(\theta)$ with an update rule 
$\theta^+=g(\theta,f(\theta))$ for the parameters $\theta$, see Figure~\ref{fig:vqa}.
Their implementation requires an iterative scheme:
For some initial parameter $\theta_0$, we evaluate $f(\theta_0)$ by 
running
the quantum algorithm~\eqref{eq:vqa_f_definition}.
As explained in ``Projective Measurement as Expectation Estimation'', evaluating the quadratic form~\eqref{eq:vqa_f_definition} requires multiple 
executions (also called \emph{shots}) of the same quantum algorithm, followed by a statistical estimation procedure.
Based on the result $f(\theta_0)$, we then compute a new parameter $\theta_1$ via the classical update rule $\theta_1=g(\theta_0,f(\theta_0))$.
This parameter is again fed into the quantum algorithm to obtain 
$f(\theta_1)$, and the iteration continues.
Due to the combination of classical and quantum elements, VQAs are often referred to as \emph{hybrid} quantum-classical algorithms.

It should be clear that VQAs are very interesting from a control perspective:
\emph{They are feedback interconnections}, consisting of a static nonlinearity and a discrete-time dynamical system!
In particular, VQAs can be written as discrete-time nonlinear systems
\begin{align}
    \theta_{k+1}&=g(\theta_k,f(\theta_k)).
\end{align}
Typically, $f$ encodes a cost function that is to be minimized.
In this case, $g$ should be designed such that it steers $\theta$ to a minimizer of the optimization problem
\begin{align}\label{eq:vqa_opt}
    \min_{\theta\in\bbr^N}f(\theta).
\end{align}
Inspired by classical optimization, a popular approach is to perform gradient descent
\begin{align}
    \theta_{k+1}=\theta_k-\gamma\nabla f(\theta_k)
\end{align}
with step-size $\gamma>0$.
A remarkable result from the recent VQA literature is that the gradient $\nabla f(\theta)$ can be \emph{exactly} determined based on evaluations of the function $f(\theta)$. 
This result is referred to as the \emph{parameter-shift rule}~\cite{bergholm2018pennylane,mitarai2018quantum,schuld2019gradient}.
Since, in practice, $f(\theta)$ cannot be retrieved exactly but is estimated by running the quantum circuit repeatedly 
and building statistics (compare~\eqref{eq:measurement_quadratic_function_estimate}),
one obtains a \emph{stochastic} gradient descent algorithm~\cite{sweke2020stochastic}.

Convergence of stochastic gradient descent for VQA optimization was studied by~\cite{sweke2020stochastic,harrow2021low} under suitable
convexity assumptions on $f$.
In general, however, the function $f$ is non-convex~\cite{huembeli2021characterizing} and, therefore, solving problem~\eqref{eq:vqa_opt} is challenging.
A challenge that is specific to the cost function of VQAs is that of \emph{barren plateaus}~\cite{mcclean2018barren}.
In the presence of a barren plateau, the partial derivatives of the cost function $f$ vanish exponentially with the number of qubits, 
which poses severe challenges to gradient-based optimization.
To be more precise, in order to estimate the function $f$ via projective measurements (and thereby its gradient, for example, via the parameter-shift rule),
the number of required circuit evaluations scales exponentially with the number of qubits, eliminating any possible advantage of VQAs
over classical algorithms.
Barren plateaus can be caused by various sources~\cite{thanasilp2022exponential}:
the parameterization of $U(\theta)$ (that is, the choice of the $H_j$'s), the measurement (that is, the choice of $\calM$), as well as noise~\cite{wang2021noise}.
There is also a direct connection of barren plateaus to quantum optimal control,
which has been used to better understand their occurrences and mitigation~\cite{larocca2022diagnosing}.
There are various further research problems in the field of VQAs, which are connected, for example, to the effect of noise,
the choice of the parameterization $U(\theta)$, as well as the development of efficient training schemes~\cite{cerezo2021variational}.

\subsection{Variational quantum eigensolver}\label{subsec:vqa_vqe}

One of the earliest VQAs was the VQE proposed by~\cite{peruzzo2014variational}.
Its goal is to determine the smallest eigenvalue $E_\rmG$ of a given Hamiltonian $H$ (that is, 
a Hermitian matrix $H=H^\dagger$).
The value $E_\rmG$ is commonly referred to as the \emph{ground state energy}, and its corresponding eigenvector $\ket{\psi_\rmG}$ as the \emph{ground state}.
Mathematically, VQE aims at finding the minimal value of the optimization problem 
\begin{align}\label{eq:vqe}
    E_\rmG=\min_{\ket{\psi}\in\bbc^{2^n},\lVert\ket{\psi}\rVert=1}\braket{\psi|H|\psi}.
\end{align}
Finding $E_\rmG$ is a problem of fundamental importance, for example, in quantum chemistry, which has motivated its initial inception as well as several applications
in this field, compare~\cite{peruzzo2014variational,kandala2017hardware,google2020hartree}.
The fact that $H$ is of size $2^n\times2^n$ and, thus, extremely large for relevant problems makes~\eqref{eq:vqe} a hard optimization problem.

VQE attempts to solve this problem by choosing a parameterized ansatz
\begin{align}
    \ket{\psi(\theta)}=U(\theta)\ket{\psi_0}
\end{align}
with a parameterized unitary matrix $U(\theta)$ as in~\eqref{eq:vqa_U_def} and some initial state $\ket{\psi}_0$.
This improves the tractability of the problem, having only $N$ trainable parameters $\theta_j$, at the cost of 
a possibly limited expressivity of the parameterization.
Note that, when choosing the observable $\calM=H$, the above parameterization directly reduces~\eqref{eq:vqe} to the VQA
optimization problem~\eqref{eq:vqa_opt}.
Key questions that arise in the context of VQE are the choice of parameterization $U(\theta)$, the classical optimization algorithm, 
as well as the mitigation of noise (see~\cite{tilly2022variational} for a recent survey).

\subsection{Quantum approximate optimization algorithm}\label{subsec:vqa_qaoa}

The QAOA is widely viewed as a promising candidate for achieving a quantum advantage in the near future~\cite{crooks2018performance,blekos2023review}.
Problems that can be tackled using the QAOA include constraint satisfaction~\cite{lin2016performance} 
and max-cut problems~\cite{wang2018quantum}, which are closely connected to computationally complex problems arising in control~\cite{blondel2000survey}.
As discussed in ``Using Quantum Computers in Control'', exploiting this connection in order to solve hard control problems on a quantum computer is a promising direction for future research.

The QAOA was proposed by~\cite{farhi2014quantum} to solve combinatorial optimization problems of the form 
\begin{align}\label{eq:qaoa_opt}
    \min_{x\in\{0,1\}^n}Q(x)
\end{align}
with some cost function $Q:\{0,1\}^m\to\bbr$.
To this end, a VQA is used with a specific cost-dependent observable $\calM$ and parameterized unitary matrix $U(\theta)$.
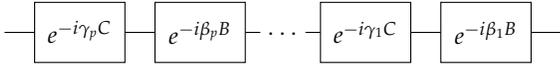
\begin{figure}
    \centerline{
        \newcommand\linesep{1}
        \begin{tikzpicture}[scale=1]
        \draw (0,0)--(0.4,0);
        \draw (0.4,0.4) rectangle (1.6,-0.4);
        \node at (1,0) {$e^{-i\gamma_pC}$};
        \draw (1.6,0)--(2,0);
        \draw (2,0.4) rectangle (3.2,-0.4);
        \node at (2.6,0) {$e^{-i\beta_pB}$};
        \draw (3.2,0)--(3.4,0);
        \node at (3.72,-0.02) {$\cdots$};
        \draw (4,0)--(4.2,0);
        \draw (4.2,0.4) rectangle (5.4,-0.4);
        \node at (4.8,0) {$e^{-i\gamma_1C}$};
        \draw (5.4,0)--(5.8,0);
        \draw (5.8,0.4) rectangle (7,-0.4);
        \node at (6.4,0) {$e^{-i\beta_1B}$};
        \draw (7,0)--(7.4,0);
        \end{tikzpicture}%
    }
    \caption{Circuit representation of the parameterized unitary matrix $U(\theta)$ used in the quantum approximate optimization algorithm (QAOA).
    The QAOA is a variational quantum algorithm which aims at solving the combinatorial optimization problem~\eqref{eq:qaoa_opt}.
    Its main distinguishing feature is the specific choice of the parameterized unitary matrix $U(\theta)$, which alternates between gates of the form $e^{-i\gamma_jC}$ and $e^{-i\beta_jB}$ for different parameters $\gamma_j,\beta_j\in\bbr$.
    The Hermitian matrices $C$ and $B$ generating these gates are carefully chosen:
    The \emph{cost Hamiltonian}
    $C$ is used to encode the cost function $Q$ from~\eqref{eq:qaoa_opt} into the quantum circuit (see~\cite{hadfield2021representation} for details), whereas the \emph{mixer Hamiltonian} $B$ is typically a sum of single-qubit Pauli-$X$ gates, compare~\eqref{eq:mixer_definition}.
    }
    \label{fig:qaoa_circuit}
\end{figure}
The ansatz chosen for $U(\theta)$ in the QAOA alternates between parameterized unitary matrices with Hermitian generators $B$ and $C$, that is,
\begin{align}\label{eq:qaoa_ansatz}
    U(\theta)=e^{-i\beta_1B}e^{-i\gamma_1C}\cdots e^{-i\beta_pB}e^{-i\gamma_pC},
\end{align}
compare Figure~\ref{fig:qaoa_circuit}.
In the generic VQA notation~\eqref{eq:vqa_U_def}--\eqref{eq:vqa_Ui_def}, this means
\begin{align}
    \theta&=(\beta_1,\gamma_1,\dots,\beta_p,\gamma_p)\in\bbr^{2p},\\
    H_{2j-1}&=B\>\>\text{and}\>\> H_{2j}=C\>\>\text{for}\>j=1,\dots,p,
\end{align}
where $N=2p$.
The two Hermitian matrices used in the QAOA are called the \emph{mixer Hamiltonian} $B$
and the \emph{cost Hamiltonian} $C$, and their meaning is as follows.
The cost Hamiltonian $C$ is chosen based on the cost function $Q(x)$ such that 
\begin{align}\label{eq:vqa_qaoa_cost_construction}
    Q(x)\ket x=C\ket x
\end{align}
holds for any $x\in\{0,1\}^n$ (see~\cite{hadfield2021representation} for an explicit procedure to construct $C$ satisfying~\eqref{eq:vqa_qaoa_cost_construction}
for a given function $Q(x)$).
The QAOA tries to solve the combinatorial optimization problem~\eqref{eq:qaoa_opt} by finding the ground state (that is, the eigenvector corresponding to the smallest eigenvalue) 
of the cost Hamiltonian $C$.
The mixer Hamiltonian, on the other hand, does not explicitly depend on the problem parameters and is typically chosen as 
a sum of single-qubit Pauli-$X$ gates, that is, 
\begin{align}\label{eq:mixer_definition}
    B&=\sum_{j=1}^n X_j
\end{align}
with $X_j$ as in~\eqref{eq:U_j_notation_definition} for $U=X$.

The initial state $\ket{\psi_0}$ of the QAOA is the ground state of the mixer Hamiltonian $B$, that is, 
according to~\eqref{eq:mixer_definition}, the state $\ket{+}^{\otimes n}$ which is in uniform superposition over all computational basis states.
By alternating between unitaries of the form $e^{-i\beta_jB}$ and $e^{-i\gamma_jC}$, the QAOA transfers the quantum state 
from the ground state of $B$ (the initial state) to the ground state of $C$ (the desired state).
In fact, one can show that the state $U(\theta)\ket{\psi_0}$ converges to the ground state of $C$ for $p\to\infty$~\cite{farhi2014quantum,farhi2000quantum}.
In practical setups, however, $p$ can typically only be chosen very small.
Therefore, in current implementations, the QAOA is mostly used as a heuristic and deriving formal guarantees for its performance is challenging.

\subsection{Quantum machine learning}\label{subsec:vqa_qml}

Quantum machine learning (QML) refers to the intersection of two scientific disciplines:
machine learning and quantum computing.
There are different possibilities for intersecting these two fields.
For example, one can employ quantum computing in order to speed up algorithms in classical machine learning~\cite{biamonte2017quantum}.
An alternative viewpoint which has gained increasing attention in recent years is to use
quantum computing directly to solve problems in supervised, unsupervised, and reinforcement learning.
We refer to~\cite{schuld2021machine,cerezo2022challenges,meyer2022survey} for recent introductions to and 
overviews over this rapidly growing research field.
In the following, we introduce a popular approach for solving supervised learning problems using VQAs.
This approach was proposed by~\cite{farhi2018classification,benedetti2019generative,schuld2020circuit} and a review can be found in~\cite{benedetti2019parameterized}.

\begin{figure}
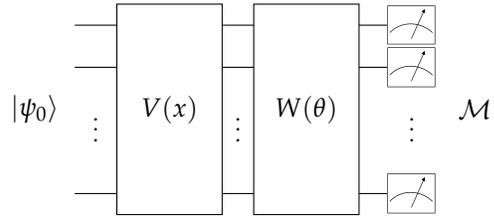

    \centerline{
        \begin{tikzpicture}[scale = 1.4]
            \draw (0,-0.2)--(0.4,-0.2);
            \draw (0,-0.6)--(0.4,-0.6);
            \draw (0,-1.8)--(0.4,-1.8);
            \node at (0.2,-1.1) {$\mathbf{\cdot}$};
            \node at (0.2,-1.2) {$\mathbf{\cdot}$};
            \node at (0.2,-1.3) {$\mathbf{\cdot}$};
            \node at (-0.38,-1) {\large $\ket{\psi_0}$};
            \draw (0.4,0) rectangle (1.4,-2);
            \node at (0.9,-1) {\large $V(x)$};
            \draw (1.4,-0.2)--(1.7,-0.2);
            \draw (1.4,-0.6)--(1.7,-0.6);
            \draw (1.4,-1.8)--(1.7,-1.8);
            \node at (1.55,-1.1) {$\mathbf{\cdot}$};
            \node at (1.55,-1.2) {$\mathbf{\cdot}$};
            \node at (1.55,-1.3) {$\mathbf{\cdot}$};
            \draw (1.7,0) rectangle (2.7,-2);
            \node at (2.2,-1) {\large $W(\theta)$};
            \draw (2.7,-0.2)--(3,-0.2);
            \draw (2.7,-0.6)--(3,-0.6);
            \draw (2.7,-1.8)--(3,-1.8);
            \node at (3.2,-1.1) {$\mathbf{\cdot}$};
            \node at (3.2,-1.2) {$\mathbf{\cdot}$};
            \node at (3.2,-1.3) {$\mathbf{\cdot}$};
            \node at (3.2,-0.2) {\includegraphics[width=0.25in]{Figures/measurement}};
            \node at (3.2,-0.6) {\includegraphics[width=0.25in]{Figures/measurement}};
            \node at (3.2,-1.8) {\includegraphics[width=0.25in]{Figures/measurement}};
            \node at (3.8,-1) {\large $\calM$};
        \end{tikzpicture}%
    }
    \caption{Circuit representation of a quantum machine learning circuit.
    The circuit consists of two main building blocks:
    The unitary matrix $V(x)$ encodes the input data into the classifier and the unitary matrix $W(\theta)$ contains the trainable part, where the parameters $\theta$ are updated in a similar fashion as for variational quantum algorithms.
    Note the ordering of $V(x)$ and $W(\theta)$ in the circuit, that is, first the data are encoded into the quantum state $V(x)\ket{\psi_0}$ and then a parameterized quantum circuit $W(\theta)$ is applied, producing $W(\theta)V(x)\ket{\psi_0}$.}
    \label{fig:qml}
\end{figure}

The basic idea is as follows.
Recall that the parameterized quantum circuit $f$ in~\eqref{eq:vqa_f_definition}
is a static nonlinear function.
In particular, for any parameter $\theta\in\bbr^N$, we can evaluate $f(\theta)$ by (repeatedly) executing a quantum algorithm.
In variational QML, $f$ is used as a function approximator, providing a new function class which is an alternative to commonly used ones such as 
neural networks or kernel models.
More precisely, consider an unknown function $\hat{f}$, for example, arising from a classification or regression task, of which we only have input-output data $\{x^i,y^i\}_{i=1}^L$ satisfying
\begin{align}
    y^i=\hat{f}(x^i)
\end{align}
for $i=1,\dots,L$ with $x^i\in\bbr^D$, $y^i\in\bbr$.
We now want to find a function $f$ of the form~\eqref{eq:vqa_f_definition} which renders the error 
\begin{align}\label{eq:qml_least_squares}
    \sum_{i=1}^L\lVert f(x^i)-y^i\rVert^2
\end{align}
small.
To this end, we introduce two distinct types of parameters into the parameterized quantum algorithm $f$:
the vector
$\theta\in\bbr^N$ contains trainable parameters, whereas the vector $x\in\bbr^D$ contains the input data to the machine learning model.
We write $f(x,\theta)$ for the parameterized quantum algorithm evaluated at $x\in\bbr^D$, $\theta\in\bbr^N$.
The goal is to determine a value of $\theta$ based on the available data such that $f(\cdot,\theta)$ is a good approximation 
of $\hat{f}(\cdot)$.
There are different possibilities for implementing the parameterization of $f$ as a quantum algorithm.
The simplest possibility is to first \emph{encode} the data into a data-dependent unitary matrix
\begin{align}
    V(x)=V_1(x_1)\cdots V_D(x_D),
\end{align}
which is followed by a trainable unitary matrix
\begin{align}
    W(\theta)=W_1(\theta_1)\cdots W_N(\theta_N).
\end{align}
The individual unitary matrices take the form
\begin{align}
    V_k(x_k)=e^{-ix_kH_{\rmV,k}},\>\>
    W_j(\theta_j)=e^{-i\theta_jH_{\rmW,j}}
\end{align}
for $k=1,\dots,D$ and $j=1,\dots,N$ with some $H_{\rmV,k}=H_{\rmV,k}^\dagger$, $H_{\rmW,j}=H_{\rmW,j}^\dagger$.
In combination, the overall variational QML circuit is defined as
\begin{align}
    f(x,\theta)=\braket{\psi_0|U(x,\theta)^\dagger\calM U(x,\theta)|\psi_0}
\end{align}
with some observable $\calM$ and the parameterized unitary matrix
\begin{align}\label{eq:qml_U_def}
    U(x,\theta)=W(\theta)V(x).
\end{align}
The circuit is schematically illustrated in Figure~\ref{fig:qml}.
Thus, in order to find a mapping $f$ minimizing the training error~\eqref{eq:qml_least_squares}, the following optimization problem 
needs to be solved 
\begin{align}\label{eq:vqa_qml_opt}
    \min_{\theta\in\bbr^N}\sum_{i=1}^L\lVert f(x^i,\theta)-y^i\rVert^2.
\end{align}
An (approximate) solution to this problem can be obtained using similar principles as for the general VQA problem~\eqref{eq:vqa_opt}.
The resulting (approximately) optimal parameter $\theta^*$ then leads to the model estimate $f(\cdot,\theta^*):\bbr^{D}\to\bbr$ which can 
be used to predict new values of the unknown function $\hat{f}$.

To summarize, VQAs can be employed as machine learning models by inserting both trainable and data-dependent unitaries, compare~\eqref{eq:qml_U_def}.
Due to their conceptual resemblance to classical neural networks, 
these models are often referred to as \emph{quantum neural networks}.
They have interesting and unique properties which have been studied in numerous recent publications, see~\cite{cerezo2022challenges} for a recent review.
Much attention has been devoted to studying the expressivity of the model $f(\theta,\cdot)$, for example, by 
rewriting it as a partial Fourier series~\cite{schuld2021effect} 
or by generalizing the circuit structure via additional data pre-processing steps~\cite{perez2020data}.
Trainability is another relevant problem which faces similar challenges as the general VQA optimization problem~\eqref{eq:vqa_opt}
(for example, barren plateaus).
Further, the literature contains various results on generalization properties of the trained model $f(\theta^*,\cdot)$ 
for unseen data points, compare\ \cite[Section II.C]{cerezo2022challenges}.
Finally, the robustness of such quantum models against input perturbations is critical~\cite{lu2020quantum} and can be analyzed based on Lipschitz bounds~\cite{berberich2023training}.
Despite these considerable advancements, it is an active and widely open research question to what extent quantum machine learning as explained above can bring advantages over classical machine learning~\cite{schuld2022is}.

\section{Density matrices}\label{sec:density_matrices}

So far, we have used state vectors $\ket{\psi}$ to describe quantum algorithms and their components.
In this section, we introduce an alternative framework for quantum computing based on the 
\emph{density matrix} (or \emph{density operator}).
While this framework is theoretically equivalent to the previous one based on state vectors, it is more useful for certain problems, 
including quantum errors and their correction which we 
discuss later in the tutorial.
The main price to pay is that density matrices are mathematically more abstract since the 
state of the system is now described via a matrix.
In the following, we provide the definition and some key properties of density matrices.
We closely follow the exposition in~\cite[Section 2.4]{nielsen2011quantum} and we refer the interested reader 
to this reference for a more in-depth treatment.

In order to define the density matrix, consider a set (typically called an \emph{ensemble}) of quantum states $\ket{\psi_i}$, $i=1,\dots, q$.
Suppose it is only known that the system is in one of these states, and we are given a set of probabilities $p_i$, $i=1,\dots,q$, such that it is in state $\ket{\psi_i}$ with probability $p_i$.
We can then define the density matrix $\rho\in\bbc^{2^n\times 2^n}$ of the system as 
\begin{align}\label{eq:density_matrix_definition}
    \rho=\sum_{i=1}^qp_i\ket{\psi_i}\bra{\psi_i}.
\end{align}
That is, $\rho$ represents the state of the quantum system 
as a weighted sum of rank-one matrices corresponding to the vectors $\ket{\psi_i}$, where the weights are given by 
the respective probabilities.

All the operations previously introduced for state vectors can be reformulated for the density matrix.
For example, composition of multiple qubits is again defined via the tensor product:
If $\rho_1$ and $\rho_2$ describe the states of two individual quantum systems, then the composite density matrix is
\begin{align}\label{eq:density_matrix_composite}
    \rho=\rho_1\otimes\rho_2.
\end{align}
As before, states of the form~\eqref{eq:density_matrix_composite} are called separable, and the full class of possible density matrices can be obtained by building linear combinations of~\eqref{eq:density_matrix_composite}, where $\rho_1$ and $\rho_2$ represent computational basis states.
Further, we have seen that quantum gates $U$ act on quantum states $\ket{\psi}$ via multiplication $U\ket{\psi}$.
In order to derive the action of $U$ on a density matrix, we apply $U$ to each of the state vectors $\ket{\psi_i}$ in the ensemble~\eqref{eq:density_matrix_definition}, leading to
\begin{align}
    \sum_{i=1}^qp_iU\ket{\psi_i}\bra{\psi_i}U^\dagger=U\sum_{i=1}^qp_i\ket{\psi_i}\bra{\psi_i}U^\dagger =U\rho U^\dagger.
\end{align}
Thus, the application of a quantum gate $U$ on a state $\rho$ is given by $\rho\mapsto U\rho U^\dagger$.
Measurements can also be described for density matrices.
For example, if we perform a projective measurement of $\rho$ with respect to the observable $\calM=\sum_{i=1}^{\ell}\lambda_iP_i$, then 
the probability of measuring $\lambda_i$ is
\begin{align}\label{eq:density_matrix_measurement}
    \rmtr(P_i\rho).
\end{align}
If the measurement result is $\lambda_i$, then, directly after the measurement, the state collapses to 
\begin{align}\label{eq:density_matrix_collapse}
    \rho_{\mathrm{after}\>\mathrm{meas.}}=
    \frac{P_i\rho P_i}{\rmtr(P_i\rho)}.
\end{align}
It is simple to verify that, if $q=1$, that is, $\rho=\ket{\psi}\bra{\psi}$ for some $\ket{\psi}$, then~\eqref{eq:density_matrix_measurement} and~\eqref{eq:density_matrix_collapse} 
are equivalent to the corresponding formulas for the state vector $\ket{\psi}$ in~\eqref{eq:measurement_probability} and~\eqref{eq:measurement_collapse}, respectively. 
States for which the density matrix takes the form $\rho=\ket{\psi}\bra{\psi}$ are called \emph{pure states}.
If the system is in a pure state, then, according to~\eqref{eq:density_matrix_definition},
 it is in the state $\ket{\psi}$ with probability $1$.
Conversely, the system is said to be in a \emph{mixed state} if $\rho$ is a non-trivial ensemble 
of multiple state vectors, that is, it is of the form~\eqref{eq:density_matrix_definition} with $q\geq 2$.
The trace of $\rho^2$ provides a simple computational criterion to distinguish between pure and mixed states:
$\rho$ is pure if and only if $\rmtr(\rho^2)=1$, and $\rmtr(\rho^2)<1$ otherwise~\cite[Exercise 2.71]{nielsen2011quantum}.

\begin{figure}
    \centerline{
        \begin{tikzpicture}[scale = 1]              
    \draw (0,0) circle (3);
    \fill (0,0) circle (0.1);
    \node at (-0.35,0.25) {\Large $z$};
    \draw[-{>[scale=2.0]}] (-3.6,0)--(3.6,0);
    \draw[-{>[scale=2.0]}] (0,3.6)--(0,-3.6);
    \node at (0.35,-3.8) {\Large $x$};
    \node at (3.8,0.35) {\Large $y$};
    \fill[color=blue] (3,0) circle (0.15);
    \fill[color=blue] (0,3) circle (0.15);
    \fill[color=blue] (0,-3) circle (0.15);
    \node at (3.5,-0.45) {{\color{blue}\Large$\ket{\psi_3}$}};
    \node at (0.5,3.5) {{\color{blue}\Large$\ket{\psi_2}$}};
    \node at (0.5,-2.5) {{\color{blue}\Large$\ket{\psi_1}$}};
    \fill[color=red] (1.2,0.6) circle (0.15);
    \node at (1.65,1) {{\color{red}\huge $\rho$}};
        \end{tikzpicture}%
    }
    \caption{Bloch sphere representation of the density matrix $\rho=\sum_{j=1}^3p_j\ket{\psi_j}\bra{\psi_j}$ (shown in red) which describes the ensemble of the pure states $\ket{\psi_1}=\frac{1}{\sqrt{2}}(\ket0+\ket1)$, $\ket{\psi_2}=\frac{1}{\sqrt{2}}(\ket0-\ket1)$, and $\ket{\psi_3}=\frac{1}{\sqrt{2}}(\ket0+i\ket1)$ (all shown in blue).
    The corresponding probabilities are $p_1=0.2$, $p_2=0.4$, $p_3=0.4$.
    The figure only shows the $x$-$y$-plane of the Bloch sphere (the $z$-components of all involved quantum states are zero).
    The coordinates of $\rho$ are a weighted linear combination of the coordinates of the $\ket{\psi_j}$'s with the respective weights given by the $p_j$'s.
    Since $\rho$ lies in the interior of the Bloch sphere, it is a mixed state, which can also be seen from $\rmtr(\rho^2)=0.6<1$.
    }
    \label{fig:density_matrix_bloch_sphere}
\end{figure}
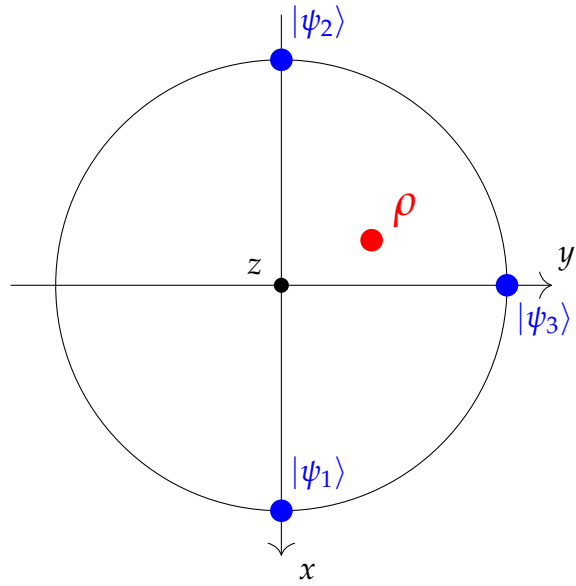

For single qubits, the difference between pure states and mixed states can be illustrated on the Bloch sphere (Figure~\ref{fig:bloch_sphere}).
Recall that the pure states are all the states on the surface of the sphere.
On the other hand, the mixed states are in the interior of the sphere, that is, in the open unit ball.
Figure~\ref{fig:density_matrix_bloch_sphere} illustrates the density matrix corresponding to a given ensemble of pure states on the Bloch sphere.

    As an illustrative example, consider a single qubit which is in state $\ket0$ or $\ket1$ with equal probability $\frac{1}{2}$, respectively.
    According to~\eqref{eq:density_matrix_definition}, the corresponding density matrix is equal to
    \begin{align}
        \rho=\frac{1}{2}\ket0\bra0+\frac{1}{2}\ket1\bra1
        =\frac{1}{2}I_2.
    \end{align}
    This state is called the \emph{maximally mixed} state since it has the 
    maximum possible uncertainty about being in any of the individual 
    state vectors.
    In the Bloch sphere (Figure~\ref{fig:bloch_sphere}), $\rho$ is in the center of the straight line 
    between the North Pole $\ket0$ and the South Pole $\ket1$, that is, at the origin.
    For a general $n$-qubit system, the maximally mixed state is $\frac{1}{2^n}I_{2^n}$.

Finally, we introduce the \emph{reduced density operator} as well as the \emph{partial trace}, which 
are useful tools for studying composite quantum systems and their subsystems.
Suppose $\rho^{AB}$ describes the joint quantum state of two systems $A$ and $B$.
The reduced density operator can be used to marginalize system $B$ and only describe the subsystem of $\rho^{AB}$ which corresponds to system $A$.
It is defined as
\begin{align}
    \rho^{A}=\rmtr_B(\rho^{AB}).
\end{align}
Here, $\rmtr_B(\rho^{AB})$ denotes the partial trace which is defined as follows:
If $\rho^{AB}=\ket{a_1}\bra{a_2}\otimes\ket{b_1}\bra{b_2}$ for pure states
$\ket{a_1}$, $\ket{a_2}$ and $\ket{b_1}$, $\ket{b_2}$ of $A$ and $B$, respectively, then
\begin{align}\label{eq:density_matrix_partial_trace_definition}
    \rmtr_B(\rho^{AB})=\ket{a_1}\bra{a_2}\rmtr(\ket{b_1}\bra{b_2}).
\end{align}
The definition of $\rmtr_B(\rho^{AB})$ is extended to arbitrary density matrices $\rho^{AB}$ by additionally requiring that 
it is a linear operator.
It is not hard to show that, if $\rho^{AB}$ is a \emph{product state}, that is, 
\begin{align}
    \rho^{AB}=\rho^A\otimes\rho^B,
\end{align}
then
\begin{align*}
    \rmtr_B(\rho^{AB})&=\rho^A,\\
    \rmtr_A(\rho^{AB})&=\rho^B.
\end{align*}
Thus, the partial trace indeed allows to retrieve a reduced state of either subsystem $A$ or $B$ by 
marginalizing the other state (compare\ \cite[Section 2.4]{nielsen2011quantum} for details).

The situation becomes more interesting when applying the partial trace to entangled states.
    Let us revisit the Bell state 
    \begin{align*}
        \ket{\Phi^+}=\frac{1}{\sqrt{2}}(\ket{00}+\ket{11})
    \end{align*}
    introduced earlier.
    As shown in~\cite[Section 2.4]{nielsen2011quantum}, the reduced density matrix of either qubit when tracing out the other one is 
    \begin{align}
        \rmtr_1(\rho)=\rmtr_2(\rho)=\frac{1}{2}I_2.
    \end{align}
    Thus, even though the two-qubit state $\ket{\Phi^+}$ is pure, each of the subsystems describing the 
    individual qubits is maximally mixed.
    This peculiar phenomenon can be attributed to the fact that $\ket{\Phi^+}$ is entangled.

\section{Errors in quantum computing}\label{sec:errors}

Noise is unavoidable in quantum computing.
It presents a key obstacle for 
reliably implementing quantum algorithms and, thus, for achieving a quantum advantage.
This was realized early on after the first proposals of quantum computing in the late 20th century,
and has stimulated substantial research efforts towards \emph{fault-tolerant quantum computing}, that is, realizing quantum computers which can work reliably in the presence of (small) errors.

In this section, we provide an introduction to errors in quantum computing.
After discussing several common error models, we introduce two main approaches that were developed to handle errors,
quantum error correction (QEC) and quantum error mitigation (QEM).
We only focus on the key ideas and refer to~\cite{nielsen2011quantum,gottesman2010introduction,devitt2013quantum,terhal2015quantum} for more detailed introductions 
to quantum errors and QEC, 
and to~\cite{cai2022quantum} for a recent survey on QEM.
Further, ``Distance Measures for Quantum States'' introduces the trace distance and the fidelity, which allow to quantify the distance between quantum states.

\begin{sidebar}{Distance Measures for Quantum States}

    \setcounter{sequation}{29}
    \renewcommand{\thesequation}{S\arabic{sequation}}
    \setcounter{stable}{0}
    \renewcommand{\thestable}{S\arabic{stable}}
    \setcounter{sfigure}{8}
    \renewcommand{\thesfigure}{S\arabic{sfigure}}

    \sdbarinitial{W}hen studying errors and their effect, we need to compare the outcome of a quantum computation on the real, noisy device 
    with the ideal outcome that would be achieved on an artificial, noise-free device.
    Thus, we need to define suitable distance measures on quantum states.    
    That this is non-trivial can be seen by considering two single-qubit pure states 
    $\ket{\psi_1}=\ket0$ and $\ket{\psi_2}=-\ket0$.
    Let us compute their norm distance
    \begin{sequation}
        \lVert\ket{\psi_1}-\ket{\psi_2}\rVert
        =\lVert 2\ket0\rVert=2.
    \end{sequation}
    On the other hand, $\ket{\psi_1}$ and $\ket{\psi_2}$ only differ by a global phase $e^{-i\pi}=-1$ and, thus, any meaningful distance 
    measure should return zero.
    Hence, the norm distance between two pure states does \emph{not} qualify as a good measure.
    
    A popular distance measure for quantum states 
    is given by the \emph{trace distance} which, for two density matrices $\rho$ and $\sigma$,
    is defined as
    \begin{sequation}
        D(\rho,\sigma)=\frac{1}{2}\rmtr(|\rho-\sigma|),
    \end{sequation}
    where $|A|=\sqrt{A^\dagger A}$.
    The trace distance has the interesting property that, for two single-qubit quantum states, 
    it is equal to $\frac{1}{2}$ times the Euclidean distance on the Bloch sphere, compare Figure~\ref{fig:bloch_sphere_trace_distance}.
    
    Another frequently employed  distance measure is provided by the \emph{fidelity}.
    The fidelity between two states $\rho$ and $\sigma$ is defined as
    \begin{sequation} 
        \calF(\rho,\sigma)=\rmtr\left(\sqrt{\sigma^{\frac{1}{2}}\rho\sigma^{\frac{1}{2}}}\right).
    \end{sequation}
    The fidelity takes values in $[0,1]$.
    It is not a usual distance in the sense that the fidelity is equal to $1$ when the two states overlap, that is, 
    $\calF(\rho,\rho)=1$, and equal to $0$ when they are orthogonal in a certain sense.
    If $\rho$ and $\sigma$ are pure, that is, $\rho=\ket{\psi_1}\bra{\psi_1}$ and $\sigma=\ket{\psi_2}\bra{\psi_2}$, 
    then the fidelity can be computed based on the more intuitive formula 
    \begin{sequation}\label{eq:fidelity_pure}
        \calF\Big(\ket{\psi_1}\bra{\psi_1},\ket{\psi_2}\bra{\psi_2}\Big)=|\braket{\psi_1|\psi_2}|.
    \end{sequation}
    For pure states $\rho=\ket{\psi_1}\bra{\psi_1}$ and $\sigma=\ket{\psi_2}\bra{\psi_2}$, the trace distance and the fidelity are connected via
    \begin{sequation}
        D\Big(\ket{\psi_1}\bra{\psi_1},\ket{\psi_2}\bra{\psi_2}\Big)=\sqrt{1-|\braket{\psi_1|\psi_2}|^2}.
    \end{sequation}

\sdbarfig{\includegraphics[width=0.7\columnwidth]{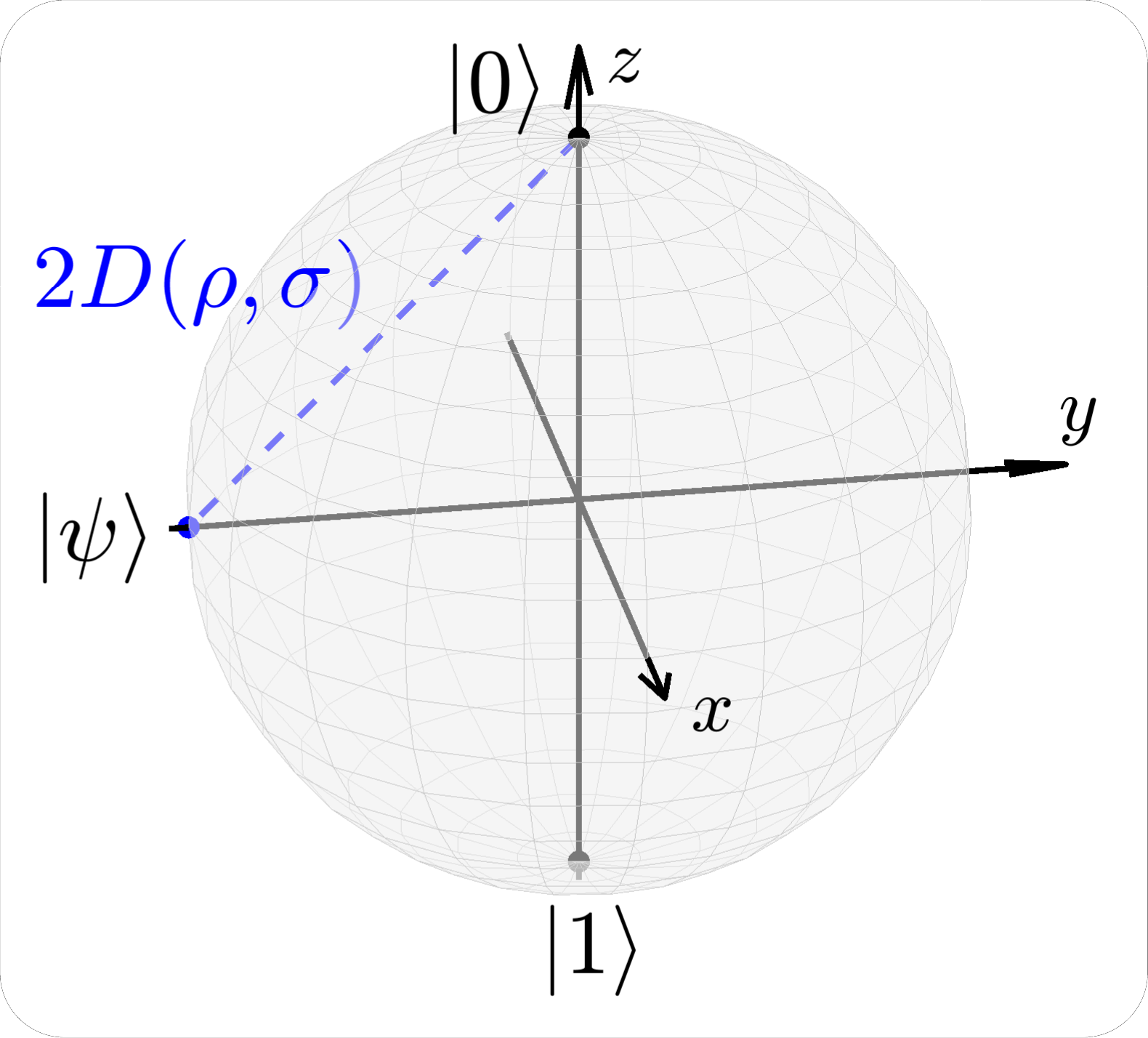}
}{Illustration of the trace distance $D(\rho,\sigma)$ between the pure states $\rho=\ket{\psi}\bra{\psi}$ (where $\ket{\psi}=\frac{1}{\sqrt{2}}(\ket0-i\ket1)$) and $\sigma=\ket0\bra0$ on the Bloch sphere, compare~\eqref{eq:bloch_sphere_angles}.
The trace distance provides a frequently used distance measure between quantum states.
For single-qubit states, it is given by $\frac{1}{2}$ times the Euclidean distance on the Bloch sphere (depicted as dashed line in the figure).\label{fig:bloch_sphere_trace_distance}}
\end{sidebar}

\subsection{Quantum errors}\label{subsec:errors_error_models}

Quantum computers face different types of errors.
For example, preparing an input state boils down to solving an 
optimal control problem subject to the Schr\"odinger 
equation~\eqref{eq:schroedinger_equation}~\cite{koch2022quantum}.
Solving this problem and implementing a suitable control input which prepares the state is in general non-trivial, especially since 
the solution typically needs to be open-loop as a measurement would collapse the state.
Any inexactness at this point can, of course, lead to a wrongly prepared input state and, thus, to an error in the overall algorithm.
Further, we have already discussed that, for many quantum algorithms, the output is a quadratic form
which can only be approximated via repeated executions of the algorithm, compare~\eqref{eq:measurement_quadratic_function_estimate}.
The resulting error is commonly referred to as \emph{shot noise}.

Even more importantly, in order to exploit quantum effects such as superposition and entanglement, qubits 
need to be kept in \emph{coherence}, that is, in perfect isolation from their environment.
Realizing this isolation over a sufficiently long time span is very challenging.
In particular, qubits are affected by \emph{decoherence} (or \emph{incoherent errors}) which describes the undesired interaction 
with the environment and the resulting loss of information~\cite{suter2016protecting}.

On the other hand, \emph{coherent errors} occur in the protected, coherent state of a qubit, and they are described by reversible, unitary operations.
For example, the implementation of a quantum gate might be inaccurate and, instead of a Pauli-$X$ rotation by a certain angle $\theta$, we might under- or 
over-rotate by $\varepsilon$ leading to an overall rotation with angle $\theta+\varepsilon$.
On the hardware level, coherent errors can be caused, for example, by inaccuracy of the control signal implementing the gate, compare\ 
\cite{kaufmann2023characterization} for a detailed characterization of coherent errors.

\begin{figure}
    \centerline{
        \newcommand\linesep{1}
        \begin{tikzpicture}[scale=1]
        \draw (0,0)--(0.4,0);
        \draw (0.4,0.4) rectangle (1.2,-0.4);
        \node at (0.8,0) {$\calE_1$};
        \draw (1.2,0)--(1.6,0);
        \draw (1.6,0.4) rectangle (2.4,-0.4);
        \node at (2,0) {$X$};
        \draw (2.4,0)--(2.8,0);
        \draw (2.8,0.4) rectangle (3.6,-0.4);
        \node at (3.2,0) {$\calE_2$};
        \draw (3.6,0)--(4,0);
        \draw (4,0.4) rectangle (4.8,-0.4);
        \node at (4.4,0) {$H$};
        \draw (4.8,0)--(5.2,0);
        \draw (5.2,0.4) rectangle (6,-0.4);
        \node at (5.6,0) {$\calE_3$};
        \draw (6,0)--(6.4,0);
        \end{tikzpicture}%
    }
    \caption{Circuit representation of quantum errors $\calE_1$, $\calE_2$, and $\calE_3$ affecting the ideal unitary matrix $U=HX$.
    Quantum errors can be classified into 1) coherent errors, which are unitary operations and are caused, for example, by imprecise gate implementations, and 2) incoherent errors, which describe the irreversible loss of information due to interaction with the environment.}
    \label{fig:error_gates}
\end{figure}
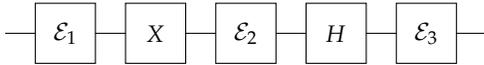

The analysis, mitigation, and correction of coherent and incoherent errors has been a central research topic in quantum computing.
On the algorithm level, these errors can be modeled as undesired operations acting on one or multiple qubits 
within the unitary matrix $U$ describing the algorithm (compare\ Figure~\ref{fig:quantum_algorithm}).
An example is illustrated in Figure~\ref{fig:error_gates}, where the ideal, error-free unitary matrix
$U=HX$ is perturbed by errors $\calE_1$, $\calE_2$, and $\calE_3$ before, in between, and after the two gates $H$ and $X$.
In the following, we provide examples of error operations resulting from both coherent and incoherent errors.

\subsubsection{Coherent errors}

An error is called coherent if it can be written as a unitary operation acting on the qubits of the algorithm.
More precisely, the error-affected state $\ket{\psi_{\mathrm{error}}}$ is equal to
\begin{align}
    \ket{\psi_{\mathrm{error}}}=
    U_{\mathrm{error}}\ket{\psi_{\mathrm{ideal}}}
\end{align}
with the ideal state $\ket{\psi_{\mathrm{ideal}}}$ and some unitary matrix $U_{\mathrm{error}}$.
For example, $U_{\mathrm{error}}$ can be a Pauli rotation $R_{\rmx}(\varepsilon)$ by a small angle $\varepsilon\in\bbr$.

An important class of coherent errors are \emph{coherent control errors}.
If the ideal quantum gate $U$ is given by $U=e^{-iH_\rmU}$ for some $H_\rmU=H_\rmU^\dagger$, then the unitary matrix
 $U_{\mathrm{error}}(\varepsilon)$ 
corresponding to a coherent control error is
\begin{align}\label{eq:cce_unitary}
    U_{\mathrm{error}}(\varepsilon)=e^{-i\varepsilon H_\rmU}
\end{align}
for some $\varepsilon\in\bbr$.
The combination of the ideal gate $U$ and the error $U_{\mathrm{error}}(\varepsilon)$ is then equal to
\begin{align}
    \tilde{U}(\varepsilon)=
    UU_{\mathrm{error}}(\varepsilon)
    =U_{\mathrm{error}}(\varepsilon)U=
    e^{-i(1+\varepsilon)H_\rmU}.
\end{align}
Here, the order of applying $U$ and $U_{\mathrm{error}}(\varepsilon)$ is irrelevant since 
$e^Ae^B=e^{A+B}$ if $A$ and $B$ commute.
Coherent control errors are an important source of error~\cite{arute2019quantum,barnes2017quantum,trout2018simulating}.
Therefore, different approaches for handling and mitigating them have been developed, for example,
composite pulses~\cite{levitt1986composite}, dynamically error-corrected gates~\cite{khodjasteh2009dynamically}, and 
randomized compiling~\cite{wallman2016noise}.

\begin{figure}
    \centerline{
    \includegraphics[width=0.8\columnwidth]{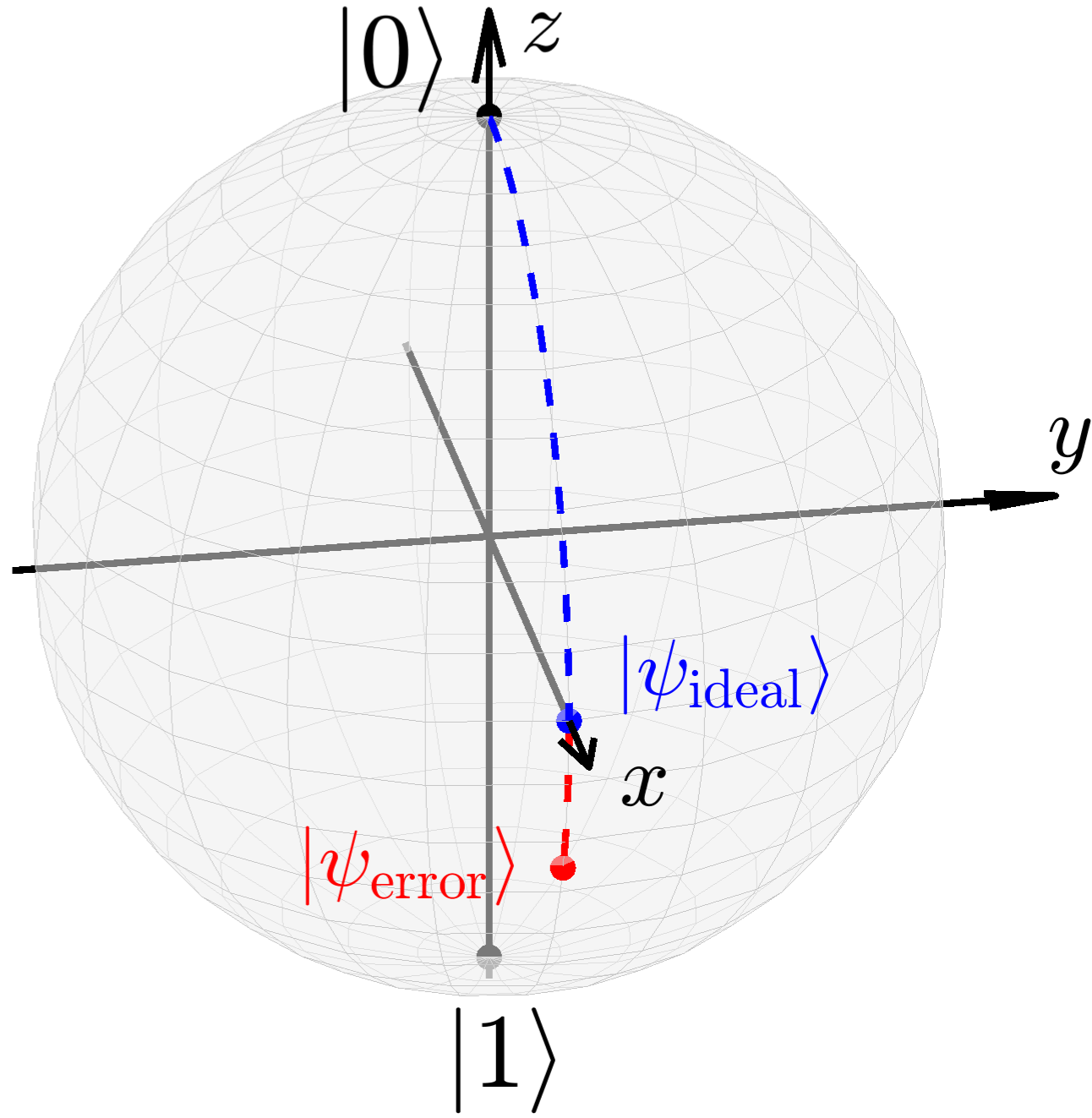}}
    \caption{Illustration of a coherent control error as an over-rotation on the Bloch sphere.
    The ideal quantum gate $U=R_{\rmy}(\theta)=e^{-i\frac{\theta}{2}Y}$ with $\theta=\frac{\pi}{2}$ is applied to the initial state $\ket0$.
    This gate is affected by a coherent control error as in~\eqref{eq:cce_unitary}, that is, an additional unitary gate $R_{\rmy}(\varepsilon)$ with $\varepsilon=\frac{\pi}{8}$ is applied to the resulting state.
    This causes an over-rotation by $25\%$.
    In the figure, the blue dashed curve depicts the evolution caused by the ideal gate (qubit states $R_{\rmy}(\theta)\ket0$ with $\theta\in[0,\frac{\pi}{2}]$), whereas the red dashed curve depicts the over-rotation (qubit states $R_{\rmy}(\frac{\pi}{2}+\varepsilon)\ket0$ with $\varepsilon\in[0,\frac{\pi}{8}]$).}
    \label{fig:bloch_sphere_Ry_cce}
\end{figure}

From a control perspective, coherent control errors are natural:
They model multiplicative errors on the Hamiltonian $H_\rmU$ implementing the quantum gate $U$.
For single-qubit gates, they can be easily interpreted on the Bloch sphere as over- or under-rotations by the 
angle $\varepsilon$, see Figure~\ref{fig:bloch_sphere_Ry_cce} for an illustration.

Using the concept of \emph{Lipschitz bounds}, it can be shown that the worst-case perturbation caused by a coherent control error 
is bounded in terms of the norm of the Hamiltonian $H_\rmU$.
To be precise, let us define
\begin{align}
    \ket{\psi}=U\ket{\psi_0}
\end{align}
for the ideal state after applying $U$ to $\ket{\psi_0}$ 
and
\begin{align}
    \ket{\tilde{\psi}(\varepsilon)}=\tilde{U}(\varepsilon)\ket{\psi_0}
\end{align}
for its noisy version.
Then, $\lVert H_\rmU\rVert$ is a Lipschitz bound for the map $\varepsilon\mapsto\ket{\tilde{\psi}(\varepsilon)}$ and, in particular, 
the fidelity (see~\eqref{eq:fidelity_pure}) between the ideal and the noisy state is bounded as 
\begin{align}\label{eq:coherent_control_errors_fidelity}
    |\braket{\tilde{\psi}(\varepsilon)|\psi}|\geq 1-\frac{\lVert H_\rmU\rVert^2\lVert \varepsilon\rVert_{\infty}^2}{2}.
\end{align}
This means that the effect of a coherent control error depends on $H_\rmU$ and, thereby, on the individual quantum gate.
This produces an interesting performance-robustness trade-off in quantum algorithms which is 
conceptually similar to classical (robust) control principles, compare\ \cite{berberich2024robustness} for details.

\subsubsection{Incoherent errors}

In the following, we discuss examples of single-qubit incoherent errors.
Incoherent errors cannot be represented by unitary operations on the qubits of the algorithm and require density matrices for their formulation.
We start with the \emph{bit flip} channel, which flips the two states $\ket0$ and $\ket1$ of a qubit 
by applying a Pauli-$X$ with probability 
$1-p$, and leaves the state unchanged otherwise.
Formally, applying the bit flip operation to an input state $\rho$ produces a noisy state 
defined by
\begin{align}\label{eq:bit_flip_operation}
    E_0\rho E_0^\dagger+E_1\rho E_1^\dagger
\end{align}
with 
\begin{align}\label{eq:bit_flip_operation_elements}
    E_0=\sqrt{p}I,\> E_1=\sqrt{1-p}X.
\end{align}
Mappings of the form~\eqref{eq:bit_flip_operation} (generally with more than two terms
$E_j\rho E_j^\dagger$) are called \emph{quantum operations}, and they 
provide a flexible and useful framework for modeling quantum errors and other operations on quantum states~\cite[Section 8]{nielsen2011quantum}.

Another incoherent error is provided by the \emph{phase flip} channel.
With probability $1-p$, it applies a Pauli-$Z$ gate, and it leaves the state unchanged otherwise.
It can also be formally described as a quantum operation~\eqref{eq:bit_flip_operation}, but with different 
matrices
\begin{align}\label{eq:phase_flip_operation_elements}
    E_0=\sqrt{p}I,\> E_1=\sqrt{1-p}Z.
\end{align}
The \emph{depolarizing channel}, on the other hand, leaves $\rho$ unchanged with probability $1-p$, and changes it to the maximally 
mixed state $\frac{1}{2}I$ otherwise.
Equivalently, applying the depolarizing channel to the input state $\rho$ produces 
\begin{align}
    \frac{p}{2}I+(1-p)\rho.
\end{align}
All incoherent errors introduced above have in common that the quantity $\rmtr(\rho^2)$ does not increase and almost always decreases.
This indicates that the quantum state after the quantum error is closer to the center of the Bloch sphere, the maximally mixed state $
\frac{1}{2}I$, implying a loss of information of the quantum state to the environment.
For a more detailed discussion of the above and further incoherent errors, we refer to~\cite[Section 8.3]{nielsen2011quantum}.

\subsection{Quantum error correction}\label{subsec:errors_QEC}

QEC is a well-developed framework and subfield of quantum computing which studies the detection and correction of quantum errors.
Conceptually, the framework is inspired by and closely related to the theory of classical error correction.
Common QEC schemes first encode the original quantum state into a possibly larger set of qubits, then perform the actual computation 
on the encoded state, and decode afterwards to detect and possibly correct errors.
However, there are several specific phenomena in quantum computing and, therefore, in QEC~\cite{nielsen2011quantum}:
First, classical errors only manifest themselves as one or multiple bit flips, whereas 
quantum errors are given by quantum operations which can map the state to a whole set of (uncountably many) values.
Further, while classical error correction can resort to redundancy, for example, transmitting multiple copies of the same bit, this is not 
easily possible for QEC due to the no-cloning theorem~\cite[Box 12.1]{nielsen2011quantum}.
Finally, measurements affect the quantum state which complicates the detection of errors.
In light of these challenges, it seems to be a miracle that QEC is possible.

Indeed, the literature contains a variety of QEC schemes for detecting and correcting quantum errors, 
see~\cite{albert2023error} for an exhaustive list.
In ``The Shor Code: Correcting Arbitrary Single-Qubit Errors'', we explain the basic idea behind the \emph{Shor code}~\cite{shor1995scheme} which protects 
qubits against \emph{arbitrary single-qubit errors}.
For more detailed introductions to the Shor code and other QEC methods, we refer to~\cite[Section 10]{nielsen2011quantum} 
as well as~\cite{gottesman2010introduction,devitt2013quantum,terhal2015quantum}.

\begin{sidebar}{The Shor Code: Correcting Arbitrary Single-Qubit Errors}

    \setcounter{sequation}{34}
    \renewcommand{\thesequation}{S\arabic{sequation}}
    \setcounter{stable}{0}
    \renewcommand{\thestable}{S\arabic{stable}}
    \setcounter{sfigure}{9}
    \renewcommand{\thesfigure}{S\arabic{sfigure}}

    \sdbarinitial{T}he literature contains a wide range of methods for performing quantum error correction (QEC).
    In the following, we introduce the basic idea behind the \emph{Shor Code}, which allows to correct arbitrary single-qubit errors.
    The Shor code combines the \emph{$3$-qubit bit flip code} and the \emph{$3$-qubit phase flip code}, which we introduce first.

    \section{The $3$-qubit bit flip code}
    
    The $3$-qubit bit flip code protects a single qubit  $\ket{\psi}=\alpha\ket0+\beta\ket1$
against the bit flip channel defined in~\eqref{eq:bit_flip_operation} and~\eqref{eq:bit_flip_operation_elements}
The qubit $\ket{\psi}$ is encoded using three qubits as 
\begin{sequation}\label{eq:QEC_code_bit_flip}
    \ket{\psi_{\mathrm{enc}}}=\alpha\ket{000}+\beta\ket{111}.
\end{sequation}
Such an encoding can, for example, be implemented by applying the gate sequence
\begin{sequation}\label{eq:QEC_bit_flip_encoding}
    (\mathrm{CNOT}\otimes I_2)(I_2\otimes \mathrm{SWAP})(\mathrm{CNOT}\otimes I_2)
\end{sequation}
to the input state  $\ket{\psi}\otimes\ket0\otimes\ket0$, compare Figure~\ref{fig:quantum_algorithm_QEC_bit_flip_encoding}.

\sdbarfig{
\begin{tikzpicture}
    \draw (0,1) -- (4,1);
    \draw (0,0) -- (4,0);    
    \draw (0,-1) -- (4,-1);
    \draw (1,1) -- (1,-0.2);
    \filldraw (1,1) circle (0.1cm);
    \draw (1,0) circle (0.2cm);
    \draw (3,1) -- (3,-0.2);
    \filldraw (3,1) circle (0.1cm);
    \draw (3,0) circle (0.2cm);       
    \draw (2,0) -- (2,-1); 
    \draw (1.8,-0.8)--(2.2,-1.2);
    \draw (1.8,-1.2)--(2.2,-0.8);
    \draw (1.8,0.2)--(2.2,-0.2);
    \draw (1.8,-0.2)--(2.2,0.2);
    \node at (-0.5,1) {\Large $\ket{\psi}$};
    \node at (-0.5,0) {\Large $\ket0$};
    \node at (-0.5,-1) {\Large $\ket0$};
    \node at (4.75,0) {\Large $\ket{\psi_{\mathrm{enc}}}$};
    \end{tikzpicture}%
    }{Quantum algorithm implementing the encoding~\eqref{eq:QEC_bit_flip_encoding} used in the $3$-qubit bit flip code.
    \label{fig:quantum_algorithm_QEC_bit_flip_encoding}}

Suppose now that each of the three qubits is affected by the bit flip channel, and a bit flip occurs for at most one qubit.
For example, if the first qubit is affected, then the state after the bit flip is equal to 
$\alpha\ket{100}+\beta\ket{011}$.
In order to detect the occurrence of a bit flip error on at most one qubit, we perform a projective measurement 
with the following four projection matrices (called \emph{error syndromes})
\begin{sequation}\label{eq:QEC_syndromes}
    P_0=\ket{000}\bra{000}+\ket{111}\bra{111},
\end{sequation}
\begin{sequation}\label{eq:QEC_syndromes2}
    P_1=\ket{100}\bra{100}+\ket{011}\bra{011},
\end{sequation}
\begin{sequation}\label{eq:QEC_syndromes3}
    P_2=\ket{010}\bra{010}+\ket{101}\bra{101},
\end{sequation}
\begin{sequation}\label{eq:QEC_syndromes4}
    P_3=\ket{001}\bra{001}+\ket{110}\bra{110},
\end{sequation}
compare~\eqref{eq:measurement_projections_decomposition}.
These correspond to measuring no error, a bit flip on qubit $1$, a bit flip on qubit $2$, and a bit flip on qubit $3$, respectively.
Depending on the outcome of the measurement, we can determine which qubit was flipped (if any).
For example, suppose that the first qubit is flipped such that the state is equal to 
\begin{sequation}\label{eq:QEC_bit_flip_state1}
    \ket{\psi_{\mathrm{flip},1}}=\alpha\ket{100}+\beta\ket{011}.
\end{sequation}
The probability for obtaining either of the four error syndromes in~\eqref{eq:QEC_syndromes}--\eqref{eq:QEC_syndromes4} is 
\begin{align}\nonumber
    \braket{\psi_{\mathrm{flip},1}|P_0|\psi_{\mathrm{flip},1}}&=0,\\\nonumber
    \braket{\psi_{\mathrm{flip},1}|P_1|\psi_{\mathrm{flip},1}}&=|\alpha|^2+|\beta|^2=1,\\\nonumber
    \braket{\psi_{\mathrm{flip},1}|P_2|\psi_{\mathrm{flip},1}}&=0,\\\nonumber
    \braket{\psi_{\mathrm{flip},1}|P_3|\psi_{\mathrm{flip},1}}&=0.
\end{align}
Thus, with probability $1$, the measurement returns the error syndrome $P_1$, from which we can conclude that the first qubit was flipped.
Hence, after the measurement, we apply the operation $X\otimes I_2\otimes I_2$ to flip the qubit back and retrieve the original state~\eqref{eq:QEC_code_bit_flip}.
Notably, the measurement only reveals information about the occurrence of an error, 
but not about the probability amplitudes $\alpha$ and $\beta$.
This is crucial in order to avoid undesired perturbations of $\alpha$ and $\beta$ due to the measurement.

\section{$3$-qubit phase flip code}
Single qubits can be protected against the phase flip error (defined in~\eqref{eq:bit_flip_operation} and~\eqref{eq:phase_flip_operation_elements}) using the $3$-qubit phase flip code.
The idea is to follow the steps of the bit flip code in a different basis.
More precisely, note that the phase flip $Z$ acts on the basis $\ket+$, $\ket-$ just like $X$ acts on the basis 
$\ket0$, $\ket1$, that is,
\begin{sequation}
    Z\ket-=\ket+,\>Z\ket+=\ket-.
\end{sequation}
Therefore, if we first perform a basis change (for example, by applying a Hadamard gate $\ket+=H\ket0$, $\ket-=H\ket1$), 
then we can follow the exact same steps as in the $3$-qubit bit flip code to detect and possibly correct any flips between $\ket+$ and $\ket-$.
More precisely, consider a qubit in the basis $\ket+$, $\ket-$, that is, 
\begin{sequation}
    \ket{\psi}=\alpha'\ket++\beta'\ket-.
\end{sequation}
Analogously to~\eqref{eq:QEC_code_bit_flip}, we can encode this state into three qubits 
\begin{sequation}
    \alpha'\ket{+++}+\beta'\ket{---}.
\end{sequation}
In order to detect the occurrence of a phase flip, we now perform a projective measurement with error syndromes 
\begin{align}\nonumber
    P_0'&=\ket{+++}\bra{+++}+\ket{---}\bra{---},\\\nonumber
    P_1'&=\ket{-++}\bra{-++}+\ket{+--}\bra{+--},\\\nonumber
    P_2'&=\ket{+-+}\bra{+-+}+\ket{-+-}\bra{-+-},\\\nonumber
    P_3'&=\ket{++-}\bra{++-}+\ket{--+}\bra{--+}.
\end{align}
As above, the measurement reveals whether a phase flip has happened (if any) and thus allows for its correction.
\end{sidebar}

\begin{sidebar}{\continuesidebar}

\setcounter{sequation}{44}
\renewcommand{\thesequation}{S\arabic{sequation}}
\setcounter{stable}{0}
\renewcommand{\thestable}{S\arabic{stable}}
\setcounter{sfigure}{0}
\renewcommand{\thesfigure}{S\arabic{sfigure}}

\section{The Shor code}
The Shor code combines the $3$-qubit bit flip code and the $3$-qubit phase flip code, 
thereby encoding a single qubit into overall $9$ qubits.
This code can detect and correct occurrences of either bit or phase flips or both at the same time.
In fact, this is enough to correct against \emph{arbitrary single-qubit errors}!
Intuitively, this remarkable fact can be explained via the collapse of the state when measuring the error syndromes.
Suppose that the original error is not precisely a bit flip or a phase flip but, for example,
a rotation $R_{\rmx}(\frac{\pi}{2})$ (half a bit flip) or just an arbitrary unitary operator.
In this case, the projective measurement still guarantees that, after the measurement, 
the state is in the image of one of the error syndromes.

In the following, we illustrate this curious phenomenon with the $3$-qubit bit flip code for simplicity, 
but 
we note that the same principle applies for the Shor code with arbitrary single-qubit errors.
Suppose that, instead of a bit flip, the quantum gate $R_{\rmx}(\frac{\pi}{2})$ is applied to the first qubit.
This produces the perturbed state 
\begin{sequation}
    \ket{\psi_{R_{\rmx}}}=\alpha \ket{-i}\otimes \ket{0}\otimes \ket{0}
    +\beta \ket{+i}\otimes\ket{1}\otimes\ket{1},
\end{sequation}
where
\begin{align*}
\ket{+i}=\frac{1}{\sqrt{2}}(\ket0+i\ket1),\>\>
\ket{-i}=\frac{1}{\sqrt{2}}(\ket0-i\ket1).
\end{align*}
When measuring this state with respect to the error syndromes~\eqref{eq:QEC_syndromes}--\eqref{eq:QEC_syndromes4}, the corresponding probabilities are 
\begin{align}\nonumber
    \braket{\psi_{R_{\rmx}}|P_0|\psi_{R_{\rmx}}}&=0.5,\>\>
    \braket{\psi_{R_{\rmx}}|P_1|\psi_{R_{\rmx}}}=0.5,\\\nonumber
    \braket{\psi_{R_{\rmx}}|P_2|\psi_{R_{\rmx}}}&=0,\quad\>
    \braket{\psi_{R_{\rmx}}|P_3|\psi_{R_{\rmx}}}=0.
\end{align}
That is, the measurement either returns $P_0$ (no bit flip) or $P_1$ (bit flip of first qubit), each with probability $50\%$.
Using~\eqref{eq:measurement_collapse}, depending on the outcome $P_0$ or $P_1$, the state after the measurement is equal to
\begin{align*}
    \frac{P_0\ket{\psi_{R_{\rmx}}}}{\sqrt{0.5}}&=\alpha\ket{000}+\beta\ket{111}\\
    \text{or}\>\>\frac{P_1\ket{\psi_{R_{\rmx}}}}{\sqrt{0.5}}&=-i(\alpha\ket{100}+\beta\ket{011}),
\end{align*}
respectively.
Hence, if $P_0$ is observed, then the state after the measurement is equal to the original state~\eqref{eq:QEC_code_bit_flip},
whereas, if $P_1$ is observed, then the state after the measurement is equal to the state 
$\ket{\psi_{\mathrm{flip},1}}$ with flipped first qubit (modulo the global phase $-i$).
Even though the error operation $R_{\rmx}(\frac{\pi}{2})$ lies in a continuum between bit flip and no bit flip,
the measurement of the error syndromes collapses the state to one of only four discrete possibilities:
no bit flip or bit flip on qubit $1$, $2$, or $3$.
Further, the measurement reveals which qubit was flipped (if any), thus allowing to correct 
the error after the measurement.

To summarize, the $3$-qubit bit flip code cannot only correct bit flips but also the error operation
$R_{\rmx}(\frac{\pi}{2})$ (as well as certain further errors).
This is made possible by the properties of projective measurements which inevitably collapse the continuum of 
possible errors onto a discrete set.
Based on the same principle, one can show that the Shor code, which combines the bit flip code and the 
phase flip code, can correct arbitrary single-qubit errors.
\end{sidebar}

Let us conclude by mentioning some important theoretical and practical aspects of QEC.
The QEC \emph{threshold theorems} guarantee the possibility of fault-tolerant quantum computing, that is, performing accurate computations despite errors,
under the assumption that the individual errors in the circuit are sufficiently small~\cite{kitaev1997quantum,aharonov1997fault,aharonov2008fault}.
Thus, assuming that quantum hardware will progress to a sufficient level,
QEC will indeed allow to implement powerful quantum algorithms with provable theoretical speedups
on actual quantum computers.
However, QEC typically produces a (possibly large) overhead in terms of additional qubits and 
gates that are required for the error detection and correction.
Therefore, in the current NISQ era, the implementation of QEC schemes is challenging and
QEC does not (yet) enable fault-tolerant quantum computing.

\subsection{Quantum error mitigation}\label{subsec:errors_QEM}

As explained above, QEC admits strong theoretical guarantees but
the limited scalability of NISQ hardware poses non-trivial challenges to its implementation.
QEM follows an alternative approach to handling errors:
Rather than adding gates for detecting and correcting errors, 
QEM post-processes the original or slightly modified algorithm on a classical computer 
in order to (partially) compensate the noise.
Although QEM faces fundamental limitations concerning the size of the errors that can be mitigated~\cite{takagi2022fundamental,quek2022exponentially},
it has proven to be useful and has, for example, played an important role in the recent demonstration of the utility of current 
quantum computers by~\cite{kim2023evidence}.

In order to provide a basic understanding of QEM, let us briefly discuss \emph{zero-noise extrapolation} (ZNE), which is 
a popular QEM method proposed by~\cite{ying2017efficient,temme2017error}.
The main idea behind ZNE is to view the output of the quantum algorithm as a function of the noise.
More precisely, the ideal algorithm provides an exact evaluation of the quadratic form $\braket{\psi_0|U^{\dagger}\calM U|\psi_0}$, compare~\eqref{eq:quantum_algorithm_quadratic_form}.
However, each of the components of the algorithm may be affected by noise $\varepsilon$:
the preparation of the initial state $\psi_0$, coherent or incoherent quantum errors affecting the unitary matrix $U$, as well as errors associated to the measurement, for example shot noise due to a statistical estimation as in~\eqref{eq:measurement_quadratic_function_estimate}.
For a fixed noise level $\bar{\varepsilon}\geq0$, running the real-world, noisy quantum computer returns a value $f(\bar{\varepsilon})$ of some (unknown) function $f$.
QEM tries to estimate the value $f(0)$ based on evaluations of $f(\bar{\varepsilon})$ for different noise levels $\bar{\varepsilon}$.
To this end, the algorithm is executed repeatedly for artificially increased noise levels $\bar{\varepsilon}$ in order to obtain a representative amount of samples from $f$.
Based on these samples, a parameterized function (for example, linear, polynomial, or exponential) is fitted.
The key idea behind ZNE is to use this fitted function in order to extrapolate to 
zero noise, thereby providing an approximation of the ideal, noise-free output of the quantum algorithm.
We refer to the recent survey~\cite{cai2022quantum} for a more detailed introduction to ZNE and further QEM methods.

\section{Conclusion}
We provided a tutorial introduction to quantum computing from the perspective of control theory.
The tutorial addressed both basic elements such as qubits, quantum gates, and measurement, as well as advanced concepts such as VQAs and quantum errors.
Throughout the tutorial, we have encountered various principles that are closely connected to control.
Performance in the sense of accuracy and efficiency is an overarching theme in quantum computing, especially proving advantages over classical algorithms, for example, in Grover's search algorithm~\cite{grover1996fast} or Shor's algorithm  for integer factorization~\cite{shor1999polynomial}.
Robustness against errors is a crucial challenge for quantum computing, especially in the NISQ era, and significant research efforts have been made in the context of QEC and QEM.
Further, scalability is an important concept in quantum computing, in particular how to design, validate, and implement algorithms with medium to large numbers of qubits and gates.
Finally, we have seen that VQAs, which are an important class of quantum algorithms, are iterative optimization algorithms or, equivalently, feedback interconnections of dynamical systems with static nonlinearities.

Quantum computing is a mathematical framework based on linear algebra and, in particular, it does not require in-depth knowledge of quantum physics.
This level of mathematical abstraction together with the occurrence of control-theoretic principles discussed above makes quantum computing an ideal application area for control theory.
Therefore, we conclude the paper with ``Research Challenges in Quantum Computing'', where we present several open challenges in the field of quantum computing with a particular emphasis on their connections to control.

\begin{sidebar}{Research Challenges in Quantum Computing}\label{sidebar_open_problems}

    \setcounter{sequation}{0}
    \renewcommand{\thesequation}{S\arabic{sequation}}
    \setcounter{stable}{0}
    \renewcommand{\thestable}{S\arabic{stable}}
    \setcounter{sfigure}{0}
    \renewcommand{\thesfigure}{S\arabic{sfigure}}

    \sdbarinitial{Q}uantum computing is a fascinating and very active research field.
In the following, we outline several important research challenges in the field of quantum computing, and we highlight their connections to control.

\subsection*{Robustness}
In the existing quantum computing literature, robustness is often viewed as a separate, secondary objective which is addressed as an additional step after the actual algorithm design, for example, via quantum error correction or quantum error mitigation.
On the other hand, it is well-known from control that robustness and performance are closely intertwined, and an ideal controller design should trade off these two objectives.
Transferring this trade-off to quantum computing is an important open problem which requires both understanding inherent robustness properties of quantum algorithms for different error classes (beyond coherent control errors as in~\eqref{eq:coherent_control_errors_fidelity}) as well as exploiting these properties during algorithm design and compilation.

\subsection*{Scalability}
A key challenge in quantum computing is the development of medium- and large-scale quantum computers, that is, devices with medium to large numbers of qubits and gates.
Quantum algorithms consist of parallel and series interconnections of quantum gates and, thus, they are inherently modular.
Exploiting this modularity to scale up quantum devices and transfer insights from noisy intermediate-scale quantum (NISQ) algorithms to larger ones provides a promising future research direction \cite{katabarwa2023early_sidebar}.
Control theory can provide a framework for addressing this challenge, being naturally suited for handling modularity by exploiting properties such as dissipativity~\cite{willems1972dissipative1_sidebar,arcak2022compositional_sidebar}.


\subsection*{Variational quantum algorithms}
From the control perspective, variational quantum algorithms (VQAs) are a very interesting class of algorithms.
They are, by definition, feedback loops and, thus, they are amenable to a wide range of tools developed in control theory.
In particular, studying systems-theoretic properties of VQAs such as (practical) stability, convergence, or robustness may lead to new theoretical guarantees on their performance and advance the understanding of NISQ algorithms in general.
Given that VQAs consist of dynamic systems interconnected with static nonlinearities, dissipativity~\cite{willems1972dissipative1_sidebar} may once again provide a useful framework, compare~\cite{lessard2022analysis_sidebar,scherer2022dissipativity_sidebar}.

\subsection*{Quantum machine learning}
Quantum machine learning (QML) is an active research field and believed to be a promising candidate for finding a quantum advantage.
Key challenges include the analysis and design of QML algorithms with a focus on trainability, expressivity, generalization, and robustness properties~\cite{cerezo2022challenges}.
Given the manifold applications of control in classical machine learning such as, for example, in the robustness of neural networks~\cite{fazlyab2019efficient_sidebar,revay2021convex_sidebar,pauli2022training_sidebar}, it would be surprising if control could not make substantial contributions to QML as well.


\subsection*{Estimation}
Numerous problems in quantum computing can be understood as estimation problems:
Quantum error mitigation~\cite{cai2022quantum} tries to filter out the influence of noise in quantum algorithms, that is, to estimate the ideal, noise-free output of the algorithm.
Quantum state tomography describes the estimation of the unknown amplitudes of a quantum state based on measurements~\cite[Section 7.7.4]{nielsen2011quantum}, whereas quantum process tomography tries to estimate quantum processes such as gates or error channels from measurements~\cite{mohseni2008quantum_sidebar}.
While various approaches have been proposed in the literature addressing these problems, they constitute active research areas with important open challenges.
Estimation methods developed in control theory may provide an alternative angle to attack these problems, possibly leading to more accurate or efficient results.

\subsection*{Applications of quantum computing}
An important challenge is to better understand the capabilities of quantum algorithms in general, but especially of available NISQ algorithms.
Using quantum algorithms in concrete practical applications may not only reveal new insights about the algorithms themselves but also lead to breakthroughs in the application area, if indeed a relevant computational problem can be solved more efficiently than before.
As explained in ``Using Quantum Computers in Control'', various computational problems relevant for control are amenable to quantum computing and, thereby, control may contribute to tackling the above challenge.


\subsection*{Theoretical guarantees via quantum control}
As explained earlier, it is highly desirable to develop quantum algorithms which admit theoretical guarantees, for example, on performance, optimality, or robustness.
Quantum control provides a framework for deriving such guarantees when controlling quantum physical systems~\cite{dong2010quantum,altafini2012modeling,dong2022quantum,koch2022quantum}.
Transferring these principles to study, improve, or develop new quantum algorithms is a promising future research field, see~\cite[Sections 5.3.3 and 5.3.4]{koch2022quantum} and~\cite{magann2021from_sidebar} for existing results in this direction.

\end{sidebar}

\begin{sidebar}{\continuesidebar}

\setcounter{sequation}{44}
\renewcommand{\thesequation}{S\arabic{sequation}}
\setcounter{stable}{0}
\renewcommand{\thestable}{S\arabic{stable}}
\setcounter{sfigure}{0}
\renewcommand{\thesfigure}{S\arabic{sfigure}}

\end{sidebar}

\section{ACKNOWLEDGMENT}
The authors are thankful to Matthias K\"ohler and Sebastian Schlor for their helpful comments, and to Timm Faulwasser for encouraging us to write this tutorial paper.
This work was funded by Deutsche Forschungsgemeinschaft
(DFG, German Research Foundation)
under Germany's Excellence Strategy - EXC
2075 - 390740016. The authors acknowledge the support
by the Stuttgart Center for Simulation Science
(SimTech).

\section{Author Information}

\begin{IEEEbiography}{{J}ulian Berberich}{\,}(julian.berberich@ist.uni-stuttgart.de) 
   received an M.Sc. degree in Engineering Cybernetics from the University of Stuttgart, Germany, in 2018. In 2022, he obtained a Ph.D. in Mechanical Engineering, also from the University of Stuttgart, Germany. He is currently working as a Lecturer (Akademischer Rat) at the Institute for Systems Theory and Automatic Control at the University of Stuttgart, Germany. In 2022, he was a visiting researcher at the ETH Zürich, Switzerland. He has received the Outstanding Student Paper Award at the 59th IEEE Conference on Decision and Control in 2020 and the 2022 George S. Axelby Outstanding Paper Award. His research interests include data-driven analysis and control as well as quantum computing.
\end{IEEEbiography}

\begin{IEEEbiography}{Daniel Fink}{\,}(daniel.fink@icp.uni-stuttgart.de)
    received an M.Sc. degree in Simulation Technology from the University of Stuttgart, Germany, in 2022. Since then, he has been working as a Doctoral Student at the Institute for Computational Physics at the University of Stuttgart. Between 2020 and 2021 he was a visiting student at the National Taiwan University (NTU). During his Ph.D. program, he is mainly working at the intersection of Quantum Computing and Machine Learning.    
\end{IEEEbiography}

\bibliographystyle{IEEEtran}
\bibliography{Literature}

\endarticle

\end{document}